\documentclass[sigconf]{acmart}
\AtBeginDocument{%
  }

\usepackage{graphicx}
\usepackage{adjustbox}
\usepackage{subcaption}
\usepackage{makecell}
\usepackage{tcolorbox}
\usepackage{pifont}
\usepackage{seqsplit}
\newcommand{\cmark}{\ding{51}}
\newcommand{\xmark}{\ding{55}}
\usepackage{pgfplots}
\pgfplotsset{compat=1.18}

\definecolor{color_graph}{HTML}{6c757d}
\usepackage{tikz}
\usetikzlibrary{shapes, arrows, positioning, fit, calc}

\usepackage{pgf-umlsd}
\usetikzlibrary{calc,fit}
\usepackage{xcolor}
\definecolor{verde}{rgb}{0.25,0.5,0.35}
\definecolor{jpurple}{rgb}{0.5,0,0.35}
\definecolor{darkgreen}{rgb}{0.0, 0.2, 0.13}
\usepackage{listings}
\usepackage{enumitem}
\usepackage[ruled,vlined]{algorithm2e}
\usepackage{microtype}

\definecolor{circl}{HTML}{E27589}

\DeclareRobustCommand{\filledcircle}[3]{%
  \tikz[baseline=(char.base)]{%
    \node[shape=circle, fill=#1, text=white, inner sep=0pt,
          minimum size=2.2ex, font=\scriptsize\bfseries] (char) {#3};%
  }%
}

\newcommand{\estiloJava}{%
  \lstset{
    language=Java,
    basicstyle=\ttfamily\scriptsize,
    keywordstyle=\color{blue},
    commentstyle=\color{gray},
    stringstyle=\color{red!70!black},
    breaklines=true,
    frame=single,
    numbers=left,
    numberstyle=\tiny\color{gray},
    showstringspaces=false,
  }%
}

\makeatletter

\newcommand{\msgarrow}[4][solid]{%
  \stepcounter{seqlevel}%
  \path (#2)+(0,-\theseqlevel*\unitfactor-0.7*\unitfactor) node (mess from) {};
  \path (#4)+(0,-\theseqlevel*\unitfactor-0.7*\unitfactor) node (mess to) {};
  \draw[->,>=angle 60,#1] (mess from) -- (mess to)
    node[midway, above, align=center] {#3};
}

\newcommand{\snote}[2]{%
  \stepcounter{seqlevel}%
  \path (#1)+(0,-\theseqlevel*\unitfactor-0.7*\unitfactor) node[
    draw, fill=yellow!20, rounded corners=2pt,
    inner sep=4pt, align=center, font=\small
  ] {#2};
  \stepcounter{seqlevel}%
}
\newcommand{\snoteover}[4][yellow!20]{%
  \stepcounter{seqlevel}%
  \node[draw, fill=#1, rounded corners=3pt,
    inner sep=8pt, align=center, font=\small]
    at ($(#2)!0.5!(#3) + (0,-\theseqlevel*\unitfactor-0.7*\unitfactor)$) {#4};
  \stepcounter{seqlevel}%
}
\newcommand{\sdgap}[1][1]{%
  \begingroup
  \edef\@oldunit{\unitfactor}%
  \def\unitfactor{#1}%
  \stepcounter{seqlevel}%
  \global\let\unitfactor\@oldunit
  \endgroup
}
\makeatother
\begin{document}
\pagestyle{plain}
\title{GapFuzz: Cross-Plane Divergence Fuzzing for Distributed SDN Controllers}

\author{Moustapha Awwalou Diouf}
\orcid{0009-0000-2063-5175}
\affiliation{%
  \institution{SnT, University of Luxembourg}
  \city{}
  \state{}
  \country{}
}
\email{moustapha.diouf@uni.lu}

\author{Samuel Ouya}
\affiliation{%
  \institution{Cheikh H. KANE Digital University}
  \city{}
  \country{}}
\email{samuel.ouya@unchk.edu.sn}

\author{Jacques Klein}
\affiliation{%
  \institution{SnT, University of Luxembourg}
  \city{}
  \country{}}
\email{jacques.klein@uni.lu}

\author{Tegawendé F. Bissyandé}
\affiliation{%
  \institution{SnT, University of Luxembourg}
  \city{}
  \country{}}
\email{tegawende.bissyande@uni.lu}

\renewcommand{\shortauthors}{M. A. Diouf et al.}

\begin{abstract}
Distributed Software-Defined Networking (SDN) clusters replicate flow state asynchronously between a master node and its backups, leaving a window during which two backup nodes can each commit a contradictory rule, the master can serialize both into the data plane, and the kernel datapath can latch onto an action that no node believes authoritative. Existing SDN fuzzers miss this fault: they confine their oracle to the control plane, target a single controller, or do not steer concurrency to provoke replication races.

We present \textsc{GapFuzz}, a stateful concurrency fuzzer for distributed SDN clusters. \textsc{GapFuzz} injects pairs of contradictory Northbound requests on two non-master nodes with controlled inter-injection delay $\Delta t$, and reconstructs the global cross-plane state by querying every replica and the kernel-datapath action through \texttt{ovs-appctl ofproto/trace}. A two-phase timing search detects whether a divergence exists, then doubles and bisects on $\Delta t$ to bound the injection-time window; a lifetime probe labels each verdict transient or persistent and assigns it to one of four cross-plane state classes derived from the ONOS~2.7 source.

On a three-node ONOS~2.7 cluster, \textsc{GapFuzz} produces a divergent verdict in 81.7\% of attempts ($N=50$, Wilson 95\% CI $[77.3, 85.4]$\%); every divergence sits between the cluster's authoritative state and the kernel datapath. Phase~2 separates a 5\,ms race window for one template from a doubling-cap regime ($\Delta t_{\max}=10.24$\,s) for six others, and 99.4\% of divergences persist past 30\,s. Replacing the kernel-datapath probe with the OpenFlow user-space probe used by prior fuzzers drops detection by 26.6 percentage points overall and by 46.5 points after excluding canonicalization-forced verdicts.

\end{abstract}





\maketitle

\section{Introduction} \label{sec:introduction}
Software-defined networking (SDN) decouples the control plane from the data plane through a logically centralized controller that programs forwarding devices via a southbound protocol such as OpenFlow~\cite{mckeown2008openflow,6994333}. Production deployments now span enterprise data centers, wide-area backbones, and cloud platforms~\cite{10.1145/2486001.2486019,10.1145/2534169.2486012}, and the controller increasingly hosts a third-party application ecosystem accessed through a Northbound REST API. That ecosystem is a primary attack surface: malicious or compromised applications have been shown to subvert the control plane, manipulate topology, poison shared state, and abuse Northbound calls in ways that traditional firewalls cannot interpose on~\cite{10.1145/3243734.3243759,10.1145/3453648,10.1145/2876019.2876024,10229058,10332465,8526819, diouf2025software}. The community has converged on the position that controller-side application threats sit at the top of the SDN risk hierarchy~\cite{10.1145/3243734.3243759,kreutz2013towards}.

As SDN scales, the control plane is no longer a single process: it is replicated across a cluster of cooperating instances that maintain a shared Network Information Base, with one node holding mastership for each switch and the others serving as backups~\cite{10.1145/2620728.2620744,8187644,10.1145/2342441.2342443}. Mainstream open-source platforms (ONOS, OpenDaylight) and operator deployments alike adopt this design. The shared state is split by purpose: cluster coordination uses consensus protocols such as Raft or Paxos~\cite{ongaro2014search,lamport2019part,sakic2019response}, while the operational state that drives forwarding---most notably the flow rule store---is replicated \emph{asynchronously} for latency reasons~\cite{SMYTH20199,8470166,6127847}. The choice is principled and well-documented, but it leaves an exploitable seam: an API call returns success after the master has committed the rule \emph{locally} and dispatched the corresponding \texttt{FlowMod} to the switch, while backups receive the update only on a later, scheduled cycle. The asynchrony of this path is the substrate of the Time-of-Check-to-Time-of-Use flaws already reported on SDN~\cite{203884,10.1145/3658644.3670301}, and it gives rise to ``Heisenbug''-style faults whose non-determinism makes them hard to reproduce with conventional testing~\cite{musuvathi2008finding}.

A growing body of work applies fuzzing to SDN~\cite{liang2018fuzzing,jero2017beads,8107437,10534286,LEE2020101720,chi2023vulnerability,287272,9681706,lee2017delta,8023155,11023293,ollando2026learning,shukla2020consistent}, but the techniques remain blind to one specific failure mode: a state in which the cluster has internally converged on rule~$A$ while the kernel datapath of the switch effectively applies rule~$B$, with no signal of the discrepancy on any control-plane interface. Single-controller fuzzers (DELTA~\cite{lee2017delta}, BEADS~\cite{jero2017beads}) cannot exercise replication races by construction. Distributed-controller fuzzers (Ambusher~\cite{10534286}) target East-West messaging but read only control-plane signals and therefore cannot witness a divergence between the cluster's view and the rule actually enforced on traffic. Cross-plane validators (AudiSDN~\cite{9681706,9155378}, CHIMERA~\cite{11023293}) compare controller state against switch state but assume a single controller and explicitly exclude the transient inconsistencies that distributed replication produces. None of the existing tools combines a data-plane oracle that reads the action effectively applied to a packet with a fuzzing strategy that drives concurrency across cluster nodes; consequently, none observes the cross-plane divergences that such concurrency produces. Section~\ref{sec:related} documents this gap quantitatively.

We close the gap with \textsc{GapFuzz}, a stateful concurrency fuzzer for distributed SDN clusters. \textsc{GapFuzz} treats the asynchronous replication window as an explicit fuzzing axis: it injects pairs of contradictory Northbound requests at two non-master nodes with a controlled inter-injection delay~$\Delta t$, then reconstructs the global state across all replicas and reads the kernel-datapath action via \texttt{ovs-appctl ofproto/trace}. A two-phase timing search first detects whether a divergence exists at $\Delta t = 0$, then doubles and bisects on $\Delta t$ to bound the injection-time window in which the divergence can still be triggered, while a lifetime probe re-observes the system after a fixed delay to separate transient races from persistent splits. The verdict is mapped onto a four-class state model derived from the ONOS~2.7 source. The result is a tool that not only detects an under-studied class of fault but also characterizes it temporally---how synchronous the injections must be, and how long the resulting divergence survives.

\smallskip
\noindent\textbf{Contributions.} This paper makes four contributions:
\begin{itemize}
    \item A \emph{root-cause analysis} of the commit-to-replicate gap in ONOS\allowbreak~2.7.0, traced through the source code of the Northbound flow installation path and reduced to three timestamps that bound the vulnerability window (\S\ref{subsec:rca}).
    \item A \emph{reproducible proof-of-concept} demonstrating an exploitable cross-plane divergence triggered through the standard Northbound REST API, with the disagreement invisible to every control-plane and OpenFlow user-space monitor (\S\ref{subsec:poc}).
    \item \emph{\textsc{GapFuzz}}, a stateful concurrency fuzzer that combines a kernel-datapath ground-truth oracle, querying the actual forwarding action via \texttt{ovs-appctl ofproto/trace}, with a two-phase temporal exploration of the inter-injection delay $\Delta t$ that characterizes both the race-condition window and the lifetime of the resulting divergence (\S\ref{sec:system_design}).
    \item An \emph{empirical evaluation} on a three-node ONOS~2.7 cluster across seven contradictory-rule templates, distinguishing race-bounded triggers from structurally persistent divergences and quantifying, through an oracle ablation, the cross-plane signal that prior tools miss (\S\ref{sec:evaluation}).
\end{itemize}

\section{Background}\label{sec:background}
\subsection{Distributed SDN Controllers}
\begin{figure}[h!]
  \centering
  \includegraphics[width=\linewidth]{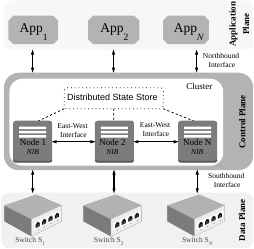}
  \caption{Architecture of a distributed SDN cluster. Each cluster node maintains a local replica of the Network Information Base (NIB) and synchronizes its shared state through a distributed state store. One node holds mastership for each switch and is alone in dispatching Southbound commands to it.}
  \label{fig:cluster-architecture}
\end{figure}

SDN architectures separate the control plane, which decides forwarding policy, from the data plane, which executes that policy on switches~\cite{6994333,mckeown2008openflow}. Applications interact with the control plane through a Northbound Interface, while the controller drives switches through a Southbound Interface, typically OpenFlow. Once a network grows beyond a single site, a single centralized controller becomes both a single point of failure and a scalability bottleneck~\cite{8187644,10.1145/2342441.2342443}. The standard remedy is to distribute the control plane across a \emph{cluster} of cooperating controller \emph{nodes}, with an East-West Interface (EWI) carrying the cooperation traffic among them. Mainstream open-source platforms (ONOS~\cite{10.1145/2620728.2620744}, OpenDaylight~\cite{8187644}) and large operator deployments such as B4~\cite{10.1145/2486001.2486019} and SWAN~\cite{10.1145/2534169.2486012} all adopt this architecture, illustrated in Figure~\ref{fig:cluster-architecture}.

Within a cluster, the shared network state---device inventory, topology, mastership, flow rules---resides in a Network Information Base (NIB) replicated across nodes. For each switch, one cluster node is elected \emph{master} and is the sole node permitted to issue Southbound commands to that switch; the remaining nodes hold backup replicas of the device's state~\cite{10.1145/2620728.2620744}. A request submitted to any node is internally routed to the master of the target device, which is responsible both for dispatching the corresponding command to the switch and for triggering replication to the backups so that the cluster's logically centralized view~\cite{10.1145/2342441.2342443} is preserved.

\subsection{State Replication and Consistency}
Replication between the master and its backups does not provide uniform guarantees across all categories of replicated state. The PACELC theorem~\cite{6127847} formalizes the trade-off: even in the absence of network partitions, a distributed system must trade latency against consistency. Distributed SDN controllers exploit this freedom by stratifying their consistency model by state type~\cite{SMYTH20199,10534286,sakic2019response}. Cluster coordination state, such as mastership election, is replicated through a consensus protocol---Raft~\cite{ongaro2014search} in ONOS, Paxos~\cite{lamport2019part} in some other deployments---requiring agreement from a quorum of nodes before any decision is acknowledged. Operational state that drives forwarding, primarily the flow rule store, is by contrast replicated \emph{asynchronously}~\cite{SMYTH20199,8470166}: the master acknowledges an operation as soon as it has been committed to its local replica, and propagates the update to backups off the acknowledgment path through a separate, scheduled mechanism.

This asynchrony leaves a temporal window during which the master's replica and the backup replicas can diverge~\cite{SMYTH20199,sakic2019response}, and no standard control-plane observation reveals it: the API has already returned success and the cluster has already routed the corresponding \texttt{FlowMod} to the switch. ONOS~\cite{10.1145/2620728.2620744} implements this model in its flow rule store, and we use ONOS as our case study throughout the paper. The same window underpins the TOCTOU vulnerabilities documented in single-controller settings~\cite{203884} and in intent-based networking~\cite{10.1145/3658644.3670301}; we instantiate it for distributed-cluster flow installation in Section~\ref{subsec:rca} and turn it into a fuzzing axis in Section~\ref{sec:system_design}.

\section{Threat Model}\label{sec:threat_model}
We consider a distributed SDN deployment in which a cluster of ONOS nodes manages one or more OpenFlow switches, with Atomix~\cite{10.1145/2620728.2620744} providing the consensus backend used by ONOS for cluster coordination, and third-party applications interacting with the cluster through the Northbound REST API. Each individual component of this stack is assumed to behave according to its specification: the ONOS source code (including its Northbound REST API authentication, the asynchronous replication mechanism of the flow rule store implemented by \texttt{FlowRuleManager} and \texttt{DeviceFlowTable}, and its mastership management); the Atomix leader-election protocol; the OpenFlow channel between the master and the switch; the East-West communication among cluster nodes; and the Open~vSwitch kernel datapath, including its revalidation mechanism~\cite{pfaff2015design}. We do not assume vulnerabilities in any of these components: the failure mode we exploit emerges from the legitimate composition of correctly behaving components, not from a defect in any one of them.

The attacker controls an authenticated third-party application that has been granted permission to install flow rules through the ONOS Northbound REST API. Such applications are routinely sourced from open-source platforms (e.g., GitHub) and integrated into operator deployments, a software supply-chain vector documented at length in the SDN security literature~\cite{10.1145/3243734.3243759,10.1145/3453648,10229058,10.1145/2876019.2876024,kreutz2013towards}. The attacker targets at least two backup nodes of the cluster and, from the compromised application, issues two concurrent flow-installation requests sharing the same selector and priority but carrying contradictory treatments. The attacker has no privileged access to the master node, the switch, the East-West channel, or the consensus backend; the attacker observes only what a legitimate Northbound client would observe.

The attacker's first goal is to drive the data plane into a forwarding state that diverges from the authoritative state held by the cluster, by exploiting the commit-to-replicate gap formalized in Section~\ref{subsec:rca}. The attacker's second goal is to keep the attack invisible to the standard control-plane and OpenFlow user-space monitoring mechanisms that an operator would use to audit the cluster: the Northbound REST API, the OpenFlow user-space flow table queried via \texttt{ovs-ofctl dump-flows}, and the cluster's role and consensus queries. Both goals must hold simultaneously: a divergence the operator can see is a bug, not an attack; an invisible state that nonetheless converges to the cluster's view is a transient anomaly, not an exploit. Section~\ref{subsec:poc} demonstrates that both goals are reachable through a single Northbound request pattern; Section~\ref{sec:system_design} turns the demonstration into a systematic fuzzing strategy.

\section{Motivation}\label{sec:motivation}
We motivate this work by establishing two facts: that the asynchronous replication path of a production distributed SDN controller mechanically produces a temporal vulnerability window during flow installation, and that this window can be turned into an exploitable, invisible cross-plane divergence through a single ordinary Northbound REST pattern. Section~\ref{subsec:rca} traces the window in the ONOS~2.7 source; Section~\ref{subsec:poc} demonstrates the resulting attack on a running cluster.

\subsection{Root Cause Analysis} \label{subsec:rca}
\begin{figure}[htbp]
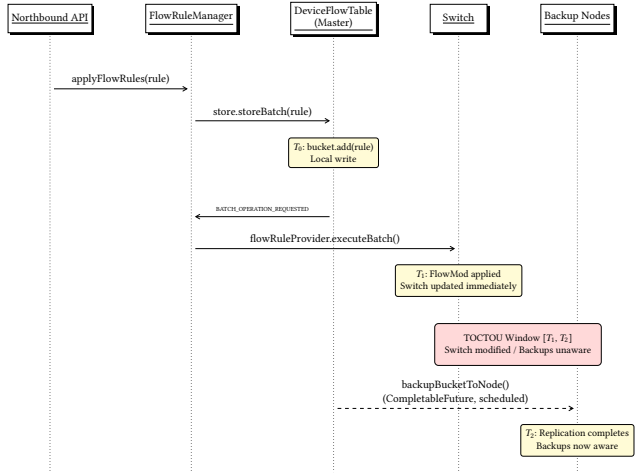

\centering
\begin{adjustbox}{scale=0.47}
\begin{sequencediagram}
\def\unitfactor{0.9}
  \newinst{api}{Northbound API}
  \newinst[1.5]{fm}{FlowRuleManager}
  \newinst[1.5]{st}{\shortstack{DeviceFlowTable\\(Master)}}
  \newinst[1.5]{sw}{Switch}
  \newinst[1.5]{bk}{Backup Nodes}
 
  \msgarrow{api}{applyFlowRules(rule)}{fm}
  \msgarrow{fm}{store.storeBatch(rule)}{st}
  \snote{st}{$T_0$: bucket.add(rule)\\Local write}
  \msgarrow{st}{\tiny{BATCH\_OPERATION\_REQUESTED}}{fm}
  \msgarrow{fm}{flowRuleProvider.executeBatch()}{sw}
  \snote{sw}{$T_1$: FlowMod applied\\Switch updated immediately}
  \snoteover[red!15]{sw}{bk}{TOCTOU Window [$T_1$, $T_2$]\\Switch modified / Backups unaware}
  \msgarrow[dashed]{st}{backupBucketToNode()\\(CompletableFuture, scheduled)}{bk}
  \snote{bk}{$T_2$: Replication completes\\Backups now aware}
  \addtocounter{seqlevel}{-1}
\end{sequencediagram}
\end{adjustbox}
\caption{Commit-to-replicate gap during flow rule installation in ONOS~2.7. The switch is updated at $T_1$, before the backup nodes are made aware of the change at $T_2$.}
\label{fig:commit-replicate-gap}
\end{figure}

We instantiate our analysis on ONOS~2.7.0\footnote{\url{https://github.com/opennetworkinglab/onos/tree/onos-2.7}} because of its industrial prevalence and because similar consistency-versus-replication trade-offs have been reported across the family of distributed SDN controllers~\cite{203884,8470166,sakic2019response}. Examining the flow rule installation path of ONOS~2.7, we sought to determine the relative ordering, on the master node, between two events: the dispatch of a \texttt{FlowMod} to the switch (data plane operation) and the propagation of the corresponding state update to backup replicas (control plane replication). The source code reveals that the two events are temporally decoupled rather than serialized through a barrier. Figure~\ref{fig:commit-replicate-gap} renders the resulting sequence; we describe each step below.

\begin{sloppypar}
When a flow rule arrives at the Northbound API, the \texttt{FlowRuleManager} delegates the operation to the distributed store via \texttt{store.storeBatch()}, which writes the rule into the local \texttt{FlowBucket} of the master node ($T_0$). That local write triggers a \texttt{BATCH\_OPERATION\_REQUESTED} event handled by \texttt{InternalStoreDelegate}, an inner class of \texttt{FlowRuleManager}, which causes the corresponding \texttt{FlowMod} to be transmitted to the switch ($T_1$); Listing~\ref{lst:flIsd} reproduces the relevant code path.
\end{sloppypar}

\begin{scriptsize}
\centering
\estiloJava
\begin{adjustbox}{max width=\linewidth}
\begin{lstlisting}[caption={\texttt{FlowRuleManager.java} -- \texttt{InternalStoreDelegate}.}, label=lst:flIsd]
private class InternalStoreDelegate implements FlowRuleStoreDelegate {
    ... 
    case BATCH_OPERATION_REQUESTED:
        // Request has been forwarded to MASTER Node
        ... 
        flowRuleProvider.executeBatch(batchOperation); 
        ... 
\end{lstlisting}
\end{adjustbox}
\end{scriptsize}
\vspace{1em}
\begin{sloppypar}
The dispatch to the switch occurs without waiting for replication to backup nodes to complete. Replication is handled by an independent, asynchronous mechanism inside the \texttt{DeviceFlowTable} ($T_2$). Listing~\ref{lst:dlbt} shows that \texttt{backupBucket()} returns a \texttt{CompletableFuture}, confirming that the replication call is non-blocking; Listing~\ref{lst:dlsc} shows that the same \texttt{backupBucket()} is invoked through \texttt{scheduler.schedule()} with a fixed \texttt{backupPeriod}, confirming that replication is delayed by up to one timer interval after $T_1$.
\vspace{1em}

\end{sloppypar}

\begin{scriptsize}
\centering
\estiloJava
\begin{adjustbox}{max width=\linewidth}
\begin{lstlisting}[caption={\texttt{DeviceFlowTable.java} -- \texttt{backupBucket()}.}, label=lst:dlbt]
private CompletableFuture<Void> backupBucket(FlowBucket bucket) {
    ... 
    if (bucket.term() == replicaInfo.term() && replicaInfo.isMaster(localNodeId)) {
        // Replicate the bucket to each of the backup nodes.
        CompletableFuture<?>[] futures = replicaInfo.backups().stream()
                .map(nodeId -> backupBucketToNode(bucket, nodeId))
                .toArray(CompletableFuture[]::new);
        return CompletableFuture.allOf(futures);
    }
    return CompletableFuture.completedFuture(null);
}
\end{lstlisting}
\end{adjustbox}
\end{scriptsize}

\begin{scriptsize}
\centering
\estiloJava
\begin{adjustbox}{max width=\linewidth}
\begin{lstlisting}[caption={\texttt{DeviceFlowTable.java} -- \texttt{scheduleBackup()}.}, label=lst:dlsc]
private void scheduleBackup(FlowBucket bucket) {
    scheduler.schedule(
            () -> executor.execute(() -> backupBucket(bucket)),
            backupPeriod, TimeUnit.MILLISECONDS);
}
\end{lstlisting}
\end{adjustbox}
\end{scriptsize}
\vspace{1em}

The interval $[T_1, T_2]$ is a concrete instantiation of the temporal vulnerability window discussed in Section~\ref{sec:introduction}: the switch's state is committed before cluster replication completes, leaving a consistency window during which the switch has been modified while the backup replicas remain unaware. A concurrent request reaching the master within this window can produce a contradictory \texttt{FlowMod} that overwrites the first one on the switch, while the distributed store may converge to a value that differs from the rule actually applied to forwarded traffic. We demonstrate this exploitation through a proof-of-concept in Section~\ref{subsec:poc}.

\subsection{Proof of Concept} \label{subsec:poc}
We turn the analysis above into a concrete, reproducible attack: a single-injection scenario that exhibits the cross-plane divergence and motivates the systematic exploration of Section~\ref{sec:system_design}. We deploy one Atomix node and three ONOS~2.7.0 controllers managing a single Open~vSwitch over OpenFlow~1.3 in a Mininet topology. From the standard Northbound REST API, we simultaneously inject two contradictory flow rules---both matching \texttt{10.0.0.99/32} at the same priority but assigning different output ports---through the REST endpoints of the two backup ONOS nodes. Both requests return \texttt{HTTP~200}; from the application's perspective, both installations have succeeded.

\begin{figure}[h!]
  \centering
  \includegraphics[width=\linewidth]{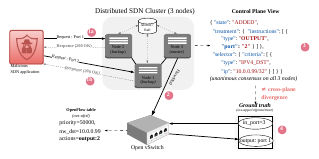}
  \caption{Cross-plane divergence captured during exploitation. Two contradictory rules are injected via backup nodes (\protect\filledcircle{circl}{black}{1a}, \protect\filledcircle{circl}{black}{1b}); the master then pushes a \texttt{FlowMod} (\protect\filledcircle{circl}{black}{2}) to the switch. The resulting state exhibits a cross-plane divergence (\protect\filledcircle{circl}{black}{3}): the control plane and the OpenFlow user-space table (\texttt{ovs-ofctl}) both hold \texttt{output:2} in unanimous consensus, while the kernel datapath---the ground truth observed via \texttt{ovs-appctl ofproto/trace} (\protect\filledcircle{circl}{black}{4})---applies \texttt{output:1}.}
  \label{fig:divergence-capture}
\end{figure}

Figure~\ref{fig:divergence-capture} captures the cluster state at $t{+}3$\,s after injection. The three ONOS nodes unanimously hold \texttt{output=port~2} in state \texttt{ADDED}, and \texttt{ovs-ofctl dump-flows} reports the same on the switch. The kernel datapath, queried via \texttt{ovs-appctl ofproto/trace}, applies \texttt{output=port~1}. The disagreement is invisible to control-plane monitoring: no API response, flow state, or cluster consensus reflects it. The OpenFlow user-space view does not reflect it either: \texttt{ovs-ofctl} agrees with the cluster while real packets follow the stale kernel megaflow~\cite{pfaff2015design}. We flush the kernel datapath cache before each run to rule out stale entries from prior tests, and we re-sample at $t{+}30$\,s: the OpenFlow table and the cluster still agree on \texttt{port~2}, while the kernel datapath still applies \texttt{port~1}. The divergence persists across the entire 30-second observation window, and we reproduced it on five independent runs, hitting at the first attempt every time.

The window $[T_1, T_2]$ of Section~\ref{subsec:rca} explains how two contradictory \texttt{FlowMod}s reach the switch in close succession through the master. Its persistence beyond that window reflects an independent data-plane reconciliation timescale---the kernel datapath revalidator of Open~vSwitch~\cite{pfaff2015design}---that no control-plane signal exposes. Detecting this class of fault requires both a data-plane oracle that inspects the action effectively applied to traffic and a fuzzing strategy that targets the temporal dynamics of distributed consensus. No existing tool combines both (Section~\ref{sec:related}); \textsc{GapFuzz} is designed to fill that gap.

\section{Cross-Plane State Model}\label{sec:model}
\subsection{Model Derivation} \label{subsec:model-derivation}
We formalize cross-plane divergence in distributed SDN clusters through a state model derived from two complementary sources. The first source is the ONOS~2.7 source code itself: the public enumerations defining flow entry states (\texttt{FlowEntry.java}) and mastership roles (\texttt{MastershipRole.java}), together with the asynchronous replication mechanism (\texttt{DeviceFlowTable.java}) traced in the root-cause analysis of Section~\ref{subsec:rca}, give us the variables that capture the cluster's distributed flow state at any moment. The second source is the implicit cross-plane contract of SDN architectures~\cite{6994333,mckeown2008openflow}: the flow state committed in the control plane should ultimately match the forwarding state enforced on the data plane, regardless of which node first acknowledged the request. While we instantiate the model on ONOS~2.7, both ingredients generalize: the variables describe any controller that exposes a per-replica flow store and a mastership role, and the contract describes any SDN architecture that distinguishes a control plane from a data plane.

\subsection{Variables and Notation} \label{subsec:state-variables}
We model the cluster state at observation time $t$, for a flow $f$ on a device $d$, as a tuple of variables observable across the cluster nodes and on the switch. We distinguish two groups of variables: those describing the \emph{control plane view} held by each of the $N$ cluster nodes, and the single variable describing the \emph{data plane view} actually enforced on the switch.

\noindent
\textbf{Control plane variables.} For each node $i \in \{1, \ldots, N\}$, three variables are observable through the management interface of node $i$:
\begin{itemize}
    \item $\text{Role}_i(d, t) \in \{\texttt{MASTER}, \texttt{STANDBY}, \texttt{NONE}\}$: the mastership role of node $i$ for device $d$ at time $t$.
    \item $\text{FlowState}_i(f, t) \in \{$ \texttt{PENDING\_ADD}, \texttt{ADDED}, \texttt{PENDING\_REMOVE}, \texttt{REMOVED}, \texttt{FAILED} $\}$: the lifecycle state of flow $f$ in the local replica of node $i$.
    \item $\text{FlowContent}_i(f, t)$: the content of flow $f$---its match criteria and forwarding action---as recorded in the local replica of node $i$.
\end{itemize}

\noindent
\textbf{Data plane variable.} On the data plane side, a single variable is observable through a forwarding-action query on the switch:
\begin{itemize}
    \item $\text{Installed\_FlowContent}(f, d, t)$: the forwarding action effectively applied to a packet matching flow $f$ at device $d$ at time $t$. This variable corresponds to the rule actually enforced on traffic, not necessarily the rule listed in the OpenFlow user-space table---the two may diverge, as our proof of concept demonstrates (Section~\ref{subsec:poc}) and as Section~\ref{sec:system_design} discusses in detail.
\end{itemize}

\noindent
\textbf{Global cross-plane state.} For a flow $f$ on a device $d$ at time $t$, the global cross-plane state combines the $N$ control plane views with the data plane view:
$$S(f, d, t) = \big(\text{CP}_1(f, d, t), \ldots, \text{CP}_N(f, d, t),\ \text{DP}(f, d, t)\big),$$
where
\begin{align*}
\text{CP}_i(f, d, t) &= \big(\text{Role}_i(d, t),\ \text{FlowState}_i(f, t),\\
&\quad \text{FlowContent}_i(f, t)\big)
\end{align*}
denotes the control plane view of node $i$, and
\begin{equation*}
\text{DP}(f, d, t) = \text{Installed\_FlowContent}(f, d, t)
\end{equation*}
denotes the data plane view.

\subsection{Invariant and Divergence} \label{subsec:invariant-classes}
For brevity we write $C_i = \text{FlowContent}_i(f, t)$, $\mathbf{C} = \{C_1, \ldots, C_N\}$, and $D = \text{Installed\_FlowContent}(f, d, t)$.

\noindent
\textbf{Cross-plane consistency invariant.} After replication completes (time $T_2$, defined in Section~\ref{subsec:rca}), every cluster node must hold the same flow content and the switch must enforce it:
$$\text{INV}:\quad \forall i, j \in \{1, \ldots, N\},\ C_i = C_j = D.$$
Observing the invariant before $T_2$ would yield false positives, since transient divergence is expected within the window $[T_1, T_2]$. In practice $T_2$ is not directly observable; the fuzzer schedules each observation after a controlled delay $\delta > T_2$ whose calibration is described in Section~\ref{sec:system_design}.

\noindent
\textbf{Divergence classes.} At any observation time $t > T_2$, the global state $S$ falls into exactly one of four mutually exclusive classes (Table~\ref{tab:divergence-matrix}):

\begin{itemize} 
    \item \textbf{CONSISTENT.} $\forall i, j: C_i = C_j = D$. The invariant holds. 
    \item \textbf{DP\_DIVERGENT.} $(\forall i, j: C_i = C_j) \land (D \notin \mathbf{C})$. The cluster has converged on a value the switch does not enforce. 
    \item \textbf{CP\_DIVERGENT.} $(\exists i, j: C_i \neq C_j) \land (D \in \mathbf{C})$. The cluster has not converged, but the switch reflects one node's view. 
    \item \textbf{FULL\_SPLIT.} $(\exists i, j: C_i \neq C_j) \land (D \notin \mathbf{C})$. The switch enforces a state held by no node. 
\end{itemize}

\begin{table}[htp]
    \centering
    \renewcommand{\arraystretch}{1.5}
    \begin{tabular}{r | c | c |}
        \multicolumn{1}{c}{} & \multicolumn{1}{c}{$D \in \mathbf{C}$} & \multicolumn{1}{c}{$D \notin \mathbf{C}$} \\
        \cline{2-3}
        $\forall i, j: C_i = C_j$ & \textbf{CONSISTENT} & \textbf{DP\_DIVERGENT} \\
        \cline{2-3}
        $\exists i, j: C_i \neq C_j$ & \textbf{CP\_DIVERGENT} & \textbf{FULL\_SPLIT} \\
        \cline{2-3}
    \end{tabular}
    \caption{Cross-plane state classes at $t > T_2$, indexed by control-plane convergence (rows) and data-plane agreement with the cluster (columns).}
    \label{tab:divergence-matrix}
\end{table}

The four-class taxonomy is a strict refinement of the binary contract \texttt{INV} that any oracle would naturally check. The two upper-row classes (\texttt{CONSISTENT} and \texttt{DP\_DIVERGENT}) correspond to a cluster that has internally agreed; the two lower-row classes capture residual disagreement among cluster replicas. Distinguishing the two columns requires a data-plane oracle that reads the action effectively applied to traffic, the gap that prior single-plane fuzzers cannot bridge by construction (Section~\ref{sec:related}).

\section{System Design}\label{sec:system_design}
\subsection{Overview and Architecture} \label{subsec:overview}
\begin{figure}[ht!]
  \centering
  \includegraphics[width=\linewidth]{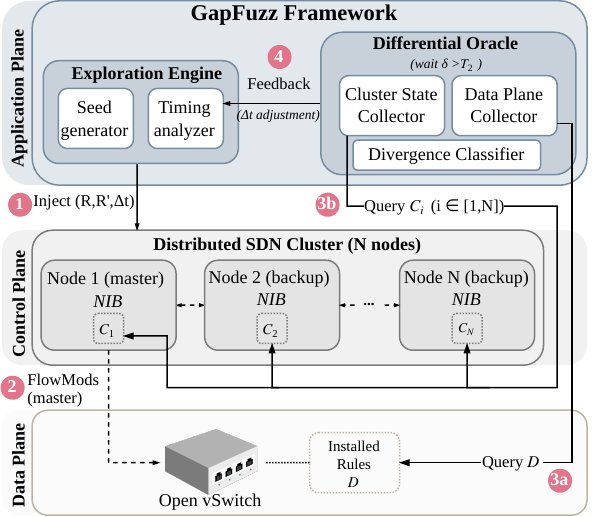}
  \caption{\textsc{GapFuzz} system overview and workflow. \protect\filledcircle{circl}{black}{1} The Exploration Engine generates contradictory pairs $(R, R')$ with timing $\Delta t$ and injects them through the cluster's Northbound REST API. \protect\filledcircle{circl}{black}{2} The master node sends the resulting \texttt{FlowMod} to the switch. The Differential Oracle then waits an observation delay $\delta > T_2$, \protect\filledcircle{circl}{black}{3a} probes the switch for the forwarding action $D$ effectively applied to a packet, and queries each cluster node for \protect\filledcircle{circl}{black}{3b} $C_i$, before classifying the cross-plane state into one of the four divergence classes. \protect\filledcircle{circl}{black}{4} The classification feeds back into the Engine to adapt $\Delta t$ for the next iteration. Solid arrows denote actions issued by \textsc{GapFuzz}; dashed arrows denote events observable in the cycle.}
  \label{fig:gapfuzz-architecture}
\end{figure}

\textsc{GapFuzz} is structured around a single design choice: every signal it consumes from the system under test is a state observable to a legitimate Northbound client or to anyone with access to the switch host, and every signal it injects is a Northbound REST request. The fuzzer thus stays inside the threat model of Section~\ref{sec:threat_model} while still observing the kernel-datapath ground truth defined in Section~\ref{subsec:state-variables}. Two components implement the loop, organized in adaptive feedback as illustrated in Figure~\ref{fig:gapfuzz-architecture}:

\begin{itemize}
    \item \textbf{Exploration Engine} (Section~\ref{subsec:exploration}): emits pairs of concurrent contradictory requests, mutating their content and the inter-injection delay $\Delta t$ to probe the asynchronous replication window $[T_1, T_2]$ identified in Section~\ref{subsec:rca}.
    \item \textbf{Differential Oracle} (Section~\ref{subsec:oracle}): reconstructs the global state $S(f, d, t)$ by querying every cluster node for its $C_i$ and the switch for $D$, then classifies it against the consistency invariant \texttt{INV} into one of the four classes of Table~\ref{tab:divergence-matrix}.
\end{itemize}

Each cycle begins with the Engine emitting a contradictory pair $(R, R')$. Once the inter-injection delay has elapsed and the wait $\delta > T_2$ has closed the replication window, the Oracle reconstructs and classifies the cross-plane state; the resulting class is fed back to the Engine, which uses it to refine $\Delta t$ for the current seed and to drive selection of subsequent seeds. Algorithm~\ref{alg:gapfuzz} (Section~\ref{subsec:algorithm}) gives the corresponding procedure.

\subsection{Exploration Engine} \label{subsec:exploration}
The Exploration Engine combines two modules---a seed generator and a timing analyzer---that together define both \emph{what} contradictory pair is sent and \emph{when} it is sent.

\noindent
\textbf{Seed generator.} The seed generator produces pairs of requests $(R, R')$ targeting the same switch, the same selector, and the same priority, but with deliberately contradictory forwarding actions. Each pair is instantiated from a parameterized YAML template; the templates we use in the evaluation cover six match fields and four actions in scope (Section~\ref{sec:eval:setup}). Generation directly emits requests acceptable to the Northbound REST API, so that no syntactic validation can short-circuit the test before it reaches the asynchronous replication path.

\noindent
\textbf{Timing analyzer.} For each seed, the engine performs a two-phase search on the inter-injection delay $\Delta t$ between $R$ and $R'$, followed by a lifetime annotation of the resulting verdict. The two phases answer two distinct questions: whether a divergence exists at all, and how synchronous the two injections must be for it to occur.

\emph{Phase~1 -- Detection.} The engine injects $R$ and $R'$ concurrently ($\Delta t = 0$), and the Oracle classifies the resulting state $S(f, d, t)$. If the verdict is \texttt{CONSISTENT}, the seed is discarded; otherwise the engine proceeds to Phase~2.

\emph{Phase~2 -- Refinement.} The engine seeks $\Delta t_{\max}$, the largest inter-injection delay for which the Oracle still returns a divergence. Starting from an initial $\Delta t_0$, the engine iteratively doubles $\Delta t$ until a \texttt{CONSISTENT} verdict is observed, then narrows the resulting interval through binary search to a fixed precision $\epsilon$. A safety cap on the number of doublings prevents unbounded growth; reaching the cap signals a divergence whose lifetime is not bounded by $\Delta t$ within the exploration budget. The empirical interpretation of the cap is discussed in Section~\ref{sec:evaluation}.

\emph{Lifetime annotation.} For each non-\texttt{CONSISTENT} verdict, the Oracle re-observes the same state $\tau_\ell$ seconds after injection (default $\tau_\ell = 30$\,s) without re-injection. The verdict is annotated \texttt{TRANSIENT} if the divergence has resolved by $\tau_\ell$ and \texttt{PERSISTENT} otherwise. The final per-seed output is the 4-tuple $(\text{seed}, \Delta t_{\max}, \text{class}, \text{lifetime})$, which separates two orthogonal temporal axes: how wide the injection-time race window is ($\Delta t_{\max}$) and how long the divergence survives once it has appeared (the lifetime label).

\subsection{Differential Oracle} \label{subsec:oracle}
\begin{sloppypar}
The Differential Oracle reconstructs the global state $S(f, d, t)$ defined in Section~\ref{subsec:state-variables} and checks it against the consistency invariant \texttt{INV}. Every observation is taken after the replication window has closed---that is, at $t > T_2$ in the notation of Section~\ref{subsec:invariant-classes}: the Oracle waits a delay $\delta > T_2$ calibrated once on the cluster's configured replication period. Three sub-modules implement the role: a Cluster State Collector, a Data Plane Collector, and a Divergence Classifier.

\noindent
\textbf{Cluster State Collector.} This sub-module queries every cluster node through the Northbound API to retrieve its control-plane view $C_i$ of flow $f$ on device $d$, including the lifecycle state and the forwarding action recorded in the local replica.

\noindent
\textbf{Data Plane Collector.} This sub-module obtains the data-plane view $D$ by probing the forwarding action effectively applied to a packet matching $f$ at device $d$. We use Open~vSwitch's \texttt{ofproto/trace} mechanism, which traces a hypothetical packet through the switch pipeline and reports the datapath actions actually computed for it---the rule effectively enforced on traffic, which may diverge from the rule listed in the OpenFlow user-space table~\cite{pfaff2015design} and from any control-plane signal (Section~\ref{subsec:poc}). The same probe could in principle be replaced by any mechanism that reads the kernel-datapath action; we use \texttt{ofproto/trace} because it is a maintained tool shipped with Open~vSwitch and does not require kernel modifications on the switch host.

\noindent
\textbf{Divergence Classifier.} Once $\mathbf{C} = \{C_1, \ldots, C_N\}$ and $D$ are collected, the classifier evaluates two binary conditions: control-plane convergence ($\forall i, j: C_i = C_j$) and data-plane agreement ($D \in \mathbf{C}$). Their combination yields the four classes of Table~\ref{tab:divergence-matrix}; the resulting class is returned to the Exploration Engine and feeds back into Phase~2 and the lifetime annotation.
\end{sloppypar}

\subsection{\textsc{GapFuzz} Algorithm} \label{subsec:algorithm}
Algorithm~\ref{alg:gapfuzz} formalizes the iterative exploration loop combining the Exploration Engine and the Differential Oracle. For each seed generated from a template, the algorithm performs the two-phase search of Section~\ref{subsec:exploration} and records the output 4-tuple. The cluster state is reset before each injection so that no residual state from previous iterations affects the current observation.

\begin{algorithm}[ht!] 
\caption{\textsc{GapFuzz} Iterative Exploration} 
\label{alg:gapfuzz} 
\KwIn{Set of templates $\mathcal{T}$, observation delay $\delta$, lifetime delay $\tau_\ell$, initial timing $\Delta t_0$, 
      precision $\epsilon$, doubling cap $D_{\max}$} 
\KwOut{Set of results $\mathcal{R}$} 
$\mathcal{R} \leftarrow \emptyset$\; 
\ForEach{template $\tau \in \mathcal{T}$}{ 
    $(R, R') \leftarrow \textsc{GenerateSeed}(\tau)$\; 
    \tcp{Phase 1 -- Detection} 
    $\textsc{Reset}()$\; 
    $\textsc{Inject}(R, R', 0)$\; 
    $\text{class} \leftarrow \textsc{Oracle}(\delta)$\; 
    \If{$\text{class} = \texttt{CONSISTENT}$}{ 
    \textbf{continue}\; 
    } 
    $\text{lifetime} \leftarrow \textsc{LifetimeCheck}(R, R', \tau_\ell)$\;
    \tcp{Phase 2 -- Refinement: exponential doubling} 
    $\Delta t \leftarrow \Delta t_0$\; 
    $\textsc{Reset}()$\; 
    $\textsc{Inject}(R, R', \Delta t)$\; 
    $\text{class} \leftarrow \textsc{Oracle}(\delta)$\; 
    $d \leftarrow 0$\;
    \While{$\text{class} \neq \texttt{CONSISTENT}$ \textbf{and} $d < D_{\max}$}{ 
        $\Delta t \leftarrow 2 \cdot \Delta t$\; 
        $d \leftarrow d + 1$\;
        $\textsc{Reset}()$\; 
        $\textsc{Inject}(R, R', \Delta t)$\; 
        $\text{class} \leftarrow \textsc{Oracle}(\delta)$\; 
    } 
    \uIf{$d = D_{\max}$ \textbf{and} $\text{class} \neq \texttt{CONSISTENT}$}{
        \tcp{cap reached: divergence not bounded by $\Delta t$}
        $\Delta t_{\max} \leftarrow \Delta t$\;
    }\Else{
        \tcp{Phase 2 -- Refinement: binary search} 
        $(\Delta t_{\max}, \text{class}) \leftarrow \textsc{BinarySearch}(R, R', \Delta t / 2, \Delta t, \epsilon, \delta)$\; 
    }
    $\mathcal{R} \leftarrow \mathcal{R} \cup \{(\tau, \Delta t_{\max}, \text{class}, \text{lifetime})\}$\; 
    } 
\Return $\mathcal{R}$\; 
\end{algorithm}

The procedure \textsc{Reset} clears, in this order, (i) cluster-side flow rules via REST DELETE on every node, (ii) the OpenFlow user-space flow table, and (iii) the kernel datapath megaflow cache. Without the third layer, megaflows that survive control-plane cleanup pollute later observations~\cite{pfaff2015design}, an artifact we describe in Section~\ref{sec:threats}. The procedure \textsc{BinarySearch} narrows the interval $[\Delta t/2, \Delta t]$ by dichotomy, invoking \textsc{Reset}, \textsc{Inject}, and \textsc{Oracle} at each midpoint until the interval width falls below the precision $\epsilon$. The returned $\Delta t_{\max}$ is the largest delay at which the Oracle still returns a divergence; \textit{class} is the class observed at $\Delta t_{\max}$.

\section{Evaluation}\label{sec:evaluation}
\subsection{Research Questions}\label{sec:eval:rq}
We evaluate \textsc{GapFuzz} against four research questions that together separate detection from temporal characterization and from oracle design:
\begin{itemize}
    \item \textbf{RQ1.} How often does \textsc{GapFuzz} observe a cross-plane divergence on a production-style distributed SDN cluster, and into which classes of the model in Section~\ref{sec:model} do those verdicts fall?
    \item \textbf{RQ2.} What is the temporal structure of these divergences along the injection-time axis: which seeds correspond to a tight write-write race ($\Delta t_{\max}$ in the millisecond range) and which correspond to a regime in which the divergence is essentially insensitive to $\Delta t$ on the doubling horizon?
    \item \textbf{RQ3.} How long do the detected divergences persist at the kernel-datapath layer (\texttt{TRANSIENT} versus \texttt{PERSISTENT}), and does the lifetime axis carry information that the injection-time axis does not?
    \item \textbf{RQ4.} Compared with a single-plane oracle of the kind that BEADS, DELTA, and Ambusher rely on, what additional cross-plane signal does the kernel-datapath oracle of \textsc{GapFuzz} contribute on the same workload?
\end{itemize}

\subsection{Implementation} \label{sec:impl}
On top of the proof-of-concept of \S\ref{subsec:poc}, \textsc{GapFuzz} adds an \texttt{asyncio} driver that executes Algorithm~\ref{alg:gapfuzz} over the templates (Phase~1, Phase~2, lifetime check), together with a Ryu companion process that exposes the ground-truth datapath state via \texttt{/dp/trace/<dpid>}. The split into two processes is forced by Ryu's \texttt{eventlet} runtime, which cannot coexist with \texttt{asyncio} in the same interpreter. Templates are YAML files consumed by \textsc{GenerateSeed}; the three-layer reset (cluster, OpenFlow user-space table, kernel megaflow cache) and the mastership-aware injection are unchanged from the proof-of-concept.

\subsection{Experimental Setup} \label{sec:eval:setup} 
\begin{sloppypar}
We run \textsc{GapFuzz} in the same environment as the proof-of-concept (\S\ref{subsec:poc}): one Atomix node, three ONOS~2.7.0 controllers, one Open~vSwitch over OpenFlow~1.3, all in a Mininet topology. Two campaigns are executed at $N=50$ each: a \emph{full} campaign (Phase~1 + Phase~2 + lifetime annotation) and a \emph{Phase-1-only} baseline (the same code path with Phase~2 disabled), so that the contribution of Phase~2 can be isolated from sampling noise. Every run cycles through the seven contradictory-flow templates listed in Table~\ref{tab:eval:corpus}; all templates lie within the match-field and action scope of the model in \S\ref{sec:model}. Algorithm parameters are fixed across campaigns at $\Delta t_0 = 10$\,ms, $\varepsilon = 1$\,ms, $\delta = 3$\,s, $\tau_\ell = 30$\,s, and $D_{\max} = 10$.
\end{sloppypar}

\begin{table}[ht!]
\centering
\small
\caption{The seven contradictory-flow templates used in the evaluation. Each row defines two rules $R$ and $R'$ that share match fields and priority but differ in their treatment.}
\label{tab:eval:corpus}
\begin{adjustbox}{valign=c, max width=\linewidth}
\begin{tabular}{l|l|l}
\toprule
Template & Match fields & Action $R$ vs Action $R'$ \\
\midrule
output\_action       & ipv4\_dst                          & OUTPUT:1 vs OUTPUT:2 \\
drop\_vs\_output     & ipv4\_dst                          & OUTPUT:1 vs DROP \\
tcp\_dst\_ports      & ipv4\_dst + ip\_proto + tcp\_dst   & OUTPUT:1 vs OUTPUT:2 \\
eth\_dst\_rewrite    & ipv4\_dst                          & SET\_FIELD eth\_dst (two MACs) \\
in\_port\_match      & ipv4\_dst + in\_port               & OUTPUT:1 vs OUTPUT:2 \\
pop\_vs\_set\_vlan   & ipv4\_dst + vlan\_vid              & POP\_VLAN vs SET\_FIELD vlan\_vid \\
set\_field\_vlan     & ipv4\_dst + vlan\_vid              & SET\_FIELD vlan\_vid (two VIDs) \\
\bottomrule
\end{tabular}
\end{adjustbox}
\end{table}

\noindent
\textbf{Metrics.} For each template $t$, the \emph{hit rate} $H_t$ is the fraction of campaign runs in which Algorithm~\ref{alg:gapfuzz} returns a divergent verdict (any class other than \texttt{CONSISTENT}); we report Wilson 95\% confidence intervals~\cite{wilson1927probable}, which behave correctly at the boundaries $H_t \to 0$ and $H_t \to 1$ where normal-approximation intervals collapse. $\Delta t_{\max}(t)$ is the value returned by Phase~2 (\S\ref{subsec:exploration}), bounded above by $\Delta t_0 \cdot 2^{D_{\max}}$ when the doubling cap is reached. The lifetime annotation classifies each divergent verdict as \texttt{TRANSIENT} (resolved by $t + \tau_{\ell}$) or \texttt{PERSISTENT} (still divergent at $t + \tau_{\ell}$).

\subsection{Results}\label{sec:eval:results}

\noindent\textbf{RQ1: How often does \textsc{GapFuzz} observe a divergence?}
Across $N=32$ full-mode campaigns---each iterating Algorithm~\ref{alg:gapfuzz} over all seven templates with a three-layer reset between templates---\textsc{GapFuzz} produces 181 divergent verdicts out of 223 attempts. The denominator is $32 \times 7 - 1 = 223$ rather than 224 because one campaign aborted before its last template completed: the companion process that proxies the data-plane oracle's \texttt{ovs-appctl ofproto/trace} invocations on the switch host (the Ryu app of \S\ref{sec:impl}) crashed mid-run, leaving \texttt{tcp\_dst\_ports} with $N=31$ instead of $N=32$. The hit rate is 81.2\% (95\% CI: 75.5--85.8\%). Every divergence falls into the \texttt{DP\_DIVERGENT} class: the kernel datapath applies an action that is absent from, or contradicts, the cluster's authoritative view. For three templates (\texttt{eth\_dst\_rewrite}, \texttt{pop\_vs\_set\_vlan}, \texttt{set\_field\_vlan}), this verdict comes from a canonicalization artifact rather than from a measured cross-plane gap; \S\ref{sec:eval:ablation} isolates the cross-plane signal on the four remaining templates. We never observe a \texttt{CP\_DIVERGENT} or \texttt{FULL\_SPLIT} verdict: the cluster nodes never disagree among themselves under our two-non-master injection pattern, leaving only the upper row of Table~\ref{tab:divergence-matrix} populated. The $N=50$ Phase-1-only baseline gives $286/350 = 81.7\%$ (95\% CI: 77.3--85.4\%), a 0.5-point gap from the full campaign that is well within sampling noise; Phase~1 alone is therefore sufficient to detect the same cross-plane gap.

\begin{figure}[ht!]
  \centering
  \adjustbox{width=\linewidth}{
  \begin{tikzpicture}
    \begin{axis}[
      xbar,
      xmode=log,
      log basis x=10,
      width=\linewidth,
      height=5.5cm,
      bar width=6pt,
      enlarge y limits=0.10,
      xmin=2e-3, xmax=1e2,
      xlabel={$\Delta t_{\max}$ (s, log scale)},
      symbolic y coords={drop,eth,inport,output,popvlan,setvlan,tcp},
      ytick=data,
      yticklabels={%
        \texttt{drop\_vs\_output},
        \texttt{eth\_dst\_rewrite},
        \texttt{in\_port\_match},
        \texttt{output\_action},
        \texttt{pop\_vs\_set\_vlan},
        \texttt{set\_field\_vlan},
        \texttt{tcp\_dst\_ports}%
      },
      yticklabel style={font=\small},
      xticklabel style={font=\small},
      xtick={1e-3,1e-2,1e-1,1,10,100},
      xticklabels={1\,ms,10\,ms,100\,ms,1\,s,10\,s,100\,s},
      grid=major,
      grid style={dotted, gray!40},
      tick align=outside,
      tick pos=left,
      legend style={
        at={(0.5,-0.32)}, anchor=north,
        legend columns=2, column sep=14pt,
        font=\footnotesize,
        draw=gray!50, fill=white,
      },
    ]
      \draw[red!70!black, dashed, thick]
        ({axis cs:10.24,drop}|-{rel axis cs:0,0}) --
        ({axis cs:10.24,drop}|-{rel axis cs:0,1});

      \addplot[xbar, fill=blue!40, draw=blue!60!black,
               nodes near coords, point meta=explicit symbolic,
               every node near coord/.append style={anchor=west,
                                                     xshift=3pt,
                                                     font=\scriptsize}]
        coordinates {
          (0.005,drop)    [$n{=}16$]
          (10.24,eth)     [$n{=}32$]
          (10.24,inport)  [$n{=}28$]
          (10.24,output)  [$n{=}15$]
          (10.24,popvlan) [$n{=}32$]
          (10.24,setvlan) [$n{=}32$]
          (10.24,tcp)     [$n{=}26$]
        };
      \addlegendentry{$\Delta t_{\max}$ per template}

      \addlegendimage{red!70!black, dashed, thick, no markers}
      \addlegendentry{doubling cap (10.24\,s)}
    \end{axis}
  \end{tikzpicture}}
  \caption{Phase~2 $\Delta t_{\max}$ per template (log scale, $N=32$ full-mode campaigns). \texttt{drop\_vs\_output} closes at 5\,ms; the other six templates reach the doubling cap at 10.24\,s, which is a lower bound on the actual injection-time window rather than a measurement. $n$ denotes the number of divergent verdicts contributing to each bar.}
  \label{fig:dtmax}
\end{figure}
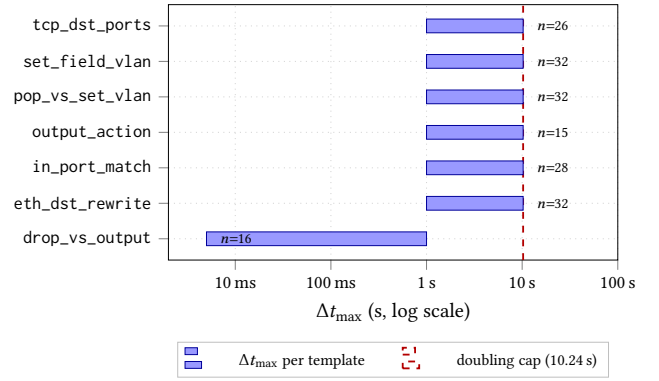

\noindent
\textbf{RQ2: What is the temporal structure of these divergences?} Phase~2 produces a $\Delta t_{\max}$ for each of the 181 divergent verdicts. The values collapse onto two points (Figure~\ref{fig:dtmax}): \texttt{drop\_vs\_output} terminates at $\Delta t_{\max} = 5$\,ms in 16/16 full-mode runs, while the other six templates reach the doubling safety cap at $\Delta t_{\max} = 10.24$\,s in 165/165 runs, with no run producing an intermediate value. The 5\,ms result is a tight write-write race that closes once concurrent edits no longer overlap; the 10.24\,s cap is an exploration-budget bound, not a measured lifetime, since Phase~2 ran out of doubling steps before observing the gap close. The cap is therefore a lower bound on the actual injection-time window for those six templates.

\noindent\textbf{RQ3: Are these divergences transient or persistent?} The lifetime annotation re-observes the cluster--datapath pair at $t + \tau_{\ell} = 30$\,s and labels each divergence \texttt{TRANSIENT} (the kernel datapath has reconciled by re-observation) or \texttt{PERSISTENT} (the gap is still present). In full mode, 180 of 181 divergences (99.4\%) are \texttt{PERSISTENT}; in the Phase-1-only baseline, the figure is 282 of 286 (98.6\%). All five \texttt{TRANSIENT} cases (one in full mode, four in the baseline) come from \texttt{drop\_vs\_output}, the same template that reaches $\Delta t_{\max} = 5$\,ms in Phase~2. The two axes therefore agree on the qualitative split: \texttt{drop\_vs\_output} hits a race window that closes within 30\,s, while the kernel-datapath state of the other six templates persists past $\tau_{\ell}$.

\noindent\textbf{Phase~1 versus Phase~2.} Phase~1 answers ``does the gap exist'' in a single shot at $\Delta t = 0$; Phase~2 answers ``how wide is the injection-time window'' by doubling and bisecting on $\Delta t$. The two campaigns let us separate these two contributions cleanly. Detection rates are statistically indistinguishable: per-template Wilson confidence intervals in Table~\ref{tab:per-template} overlap on every row, and the overall intervals overlap (full: $[75.5, 85.8]$\%; baseline: $[77.3, 85.4]$\%); Phase~2 adds no detection power on this workload. What it adds is the temporal characterization. Phase~1's lifetime probe already separates \texttt{drop\_vs\_output} (a mix of \texttt{TRANSIENT} and \texttt{PERSISTENT}) from \texttt{eth\_dst\_rewrite} (uniformly \texttt{PERSISTENT}), but only on a discrete two-label axis; Phase~2's $\Delta t_{\max}$ places them three orders of magnitude apart on a continuous one (5\,ms versus 10.24\,s). For a yes/no detection signal, Phase~1 suffices; for the full temporal picture, both axes are needed---neither subsumes the other.

\begin{table*}[ht!]
  \centering
  \caption{Per-template results across both campaigns. F = full mode (Algorithm~\ref{alg:gapfuzz} with Phase~1 + Phase~2; $N=32$ campaigns, except \texttt{tcp\_dst\_ports} where one campaign aborted before its last template, giving $N=31$). B = Phase-1-only baseline ($N=50$). Hit rate = \texttt{DP\_DIVERGENT} verdicts / attempts, with Wilson 95\% confidence intervals. Lifetime is annotated at $t + \tau_{\ell} = 30$\,s; P = \texttt{PERSISTENT}, T = \texttt{TRANSIENT}.}
  \label{tab:per-template}
  \small
  \setlength{\tabcolsep}{4pt}
  \begin{tabular}{l rcrl rcl}
    \toprule
    & \multicolumn{4}{c}{Full mode (Phase~1 + Phase~2)}
    & \multicolumn{3}{c}{Phase-1-only baseline}\\
    \cmidrule(lr){2-5}\cmidrule(lr){6-8}
    Template
      & hits/$n$ & 95\% CI & $\Delta t_{\max}$ & lifetime
      & hits/$n$ & 95\% CI & lifetime\\
    \midrule
    \texttt{drop\_vs\_output}    & 16/32 (50.0\%)  & [33.6,\,66.4] & 5\,ms     & 15\,P\,/\,1\,T & 22/50 (44.0\%) & [31.2,\,57.7] & 18\,P\,/\,4\,T \\
    \texttt{eth\_dst\_rewrite}   & 32/32 (100\%)   & [89.3,\,100]  & 10.24\,s  & 32\,P          & 50/50 (100\%)  & [92.9,\,100]  & 50\,P          \\
    \texttt{in\_port\_match}     & 28/32 (87.5\%)  & [71.9,\,95.0] & 10.24\,s  & 28\,P          & 43/50 (86.0\%) & [73.8,\,93.0] & 43\,P          \\
    \texttt{output\_action}      & 15/32 (46.9\%)  & [30.9,\,63.6] & 10.24\,s  & 15\,P          & 27/50 (54.0\%) & [40.4,\,67.0] & 27\,P          \\
    \texttt{pop\_vs\_set\_vlan}  & 32/32 (100\%)   & [89.3,\,100]  & 10.24\,s  & 32\,P          & 50/50 (100\%)  & [92.9,\,100]  & 50\,P          \\
    \texttt{set\_field\_vlan}    & 32/32 (100\%)   & [89.3,\,100]  & 10.24\,s  & 32\,P          & 50/50 (100\%)  & [92.9,\,100]  & 50\,P          \\
    \texttt{tcp\_dst\_ports}     & 26/31 (83.9\%)  & [67.4,\,92.9] & 10.24\,s  & 26\,P          & 44/50 (88.0\%) & [76.2,\,94.4] & 44\,P          \\
    \midrule
    \textbf{Overall}
      & \textbf{181/223 (81.2\%)} & \textbf{[75.5,\,85.8]} & --- & 180\,P\,/\,1\,T
      & \textbf{286/350 (81.7\%)} & \textbf{[77.3,\,85.4]} & 282\,P\,/\,4\,T \\
    \bottomrule
  \end{tabular}
\end{table*}

\subsection{Oracle ablation: native vs.\ user-space} \label{sec:eval:ablation}
\begin{sloppypar}
Research questions RQ1--RQ3 all rely on the kernel-datapath oracle; RQ4 asks what that oracle contributes over the OpenFlow user-space probe used by prior fuzzers (BEADS, DELTA, Ambusher). To answer it, we swap only the data-plane probe and re-run the $N=50$ Phase-1-only baseline. The native variant uses \texttt{ovs-appctl ofproto/trace}, which reports the effective per-packet action computed by the kernel datapath~\cite{pfaff2015design}; the user-space variant reads the switch's OpenFlow flow table via \texttt{ovs-ofctl dump-flows}. Everything else is identical: cluster collector, injection schedule, three-layer reset, templates, Phase~1. Since BEADS, DELTA, and Ambusher all observe at the controller or OpenFlow user-space level, the user-space variant's detection rate is an upper bound on what those tools could surface on this workload.

\noindent\textbf{Hit rate.} The native oracle hits in 286/350 attempts (81.7\%, 95\% CI: 77.3--85.4\%); the user-space oracle hits in 193/350 attempts (55.1\%, 95\% CI: 49.9--60.3\%). The two confidence intervals are disjoint, so the 26.6-point overall drop is not a sampling artifact.

\noindent\textbf{Per-template.} The drop concentrates on four templates (Table~\ref{tab:ablation}). For \texttt{in\_port\_match} and \texttt{tcp\_dst\_ports}, hit rate falls by 72 and 68 points respectively, with disjoint confidence intervals in both cases; for \texttt{drop\_vs\_output} and \texttt{output\_action}, by 30 and 16 points. The other three templates (\texttt{eth\_dst\_rewrite}, \texttt{pop\_vs\_set\_vlan}, \texttt{set\_field\_vlan}) report 100\% under both oracles for an artifactual reason: their actions are SET\_FIELD or POP\_VLAN, which \textsc{GapFuzz}'s canonical action set $\{\texttt{DROP}, \texttt{OUTPUT}(\text{port})\}$ cannot represent, so \texttt{DP\_DIVERGENT} fires mechanically and the comparison loses meaning on those rows.

\noindent\textbf{Lifetime.} Under the user-space oracle, 24 of 193 divergences (12.4\%) are \texttt{TRANSIENT}, against 4 of 286 (1.4\%) under the native oracle. The user-space \texttt{TRANSIENT} cases concentrate on three of the four templates where hit rate dropped (7 on \texttt{in\_port\_match}, 10 on \texttt{tcp\_dst\_ports}, 7 on \texttt{drop\_vs\_output}); the fourth (\texttt{output\_action}) reports all 19 of its user-space hits as \texttt{PERSISTENT}. The user-space view picks up disagreements that close within $\tau_{\ell}$, while the kernel-datapath probe sees a state that does not---a direct consequence of the staleness behavior of the kernel megaflow cache relative to the OpenFlow user-space table~\cite{pfaff2015design}.

\noindent\textbf{Cross-plane signal.} Removing the three artifactual templates, the native oracle detects $136/200 = 68.0\%$ against the user-space oracle's $43/200 = 21.5\%$. The remaining 46.5-point gap is a lower bound on what an oracle without access to the effective per-packet action misses on this workload---in absolute terms, 93 cross-plane divergences invisible to a controller- or OpenFlow-level oracle.
\end{sloppypar}

\begin{table*}[ht!]
  \centering
  \caption{Oracle ablation per template ($N=50$ Phase-1-only baseline; same workload; all components other than the data-plane probe are shared between the two variants). Hit rate = \texttt{DP\_DIVERGENT} verdicts / attempts, with Wilson 95\% confidence intervals. $\Delta$ is the absolute drop in hit rate from the native to the user-space oracle, in percentage points. The three templates with $\Delta = 0$ produce a tied 100\% under both oracles for an action-representation reason discussed in the text.}
  \label{tab:ablation}
  \small
  \setlength{\tabcolsep}{4pt}
  \begin{tabular}{lccccc}
    \toprule
    Template
      & \multicolumn{2}{c}{Native oracle}
      & \multicolumn{2}{c}{User-space oracle}
      & $\Delta$ \\
    \cmidrule(lr){2-3}\cmidrule(lr){4-5}
      & hits/$n$ & 95\% CI & hits/$n$ & 95\% CI & (pts) \\
    \midrule
    \texttt{drop\_vs\_output}    & 22/50 (44.0\%) & [31.2,\,57.7] & 7/50 (14.0\%)  & [7.0,\,26.2]  & 30.0 \\
    \texttt{eth\_dst\_rewrite}   & 50/50 (100\%)  & [92.9,\,100]  & 50/50 (100\%)  & [92.9,\,100]  & 0    \\
    \texttt{in\_port\_match}     & 43/50 (86.0\%) & [73.8,\,93.0] & 7/50 (14.0\%)  & [7.0,\,26.2]  & 72.0 \\
    \texttt{output\_action}      & 27/50 (54.0\%) & [40.4,\,67.0] & 19/50 (38.0\%) & [25.9,\,51.8] & 16.0 \\
    \texttt{pop\_vs\_set\_vlan}  & 50/50 (100\%)  & [92.9,\,100]  & 50/50 (100\%)  & [92.9,\,100]  & 0    \\
    \texttt{set\_field\_vlan}    & 50/50 (100\%)  & [92.9,\,100]  & 50/50 (100\%)  & [92.9,\,100]  & 0    \\
    \texttt{tcp\_dst\_ports}     & 44/50 (88.0\%) & [76.2,\,94.4] & 10/50 (20.0\%) & [11.2,\,33.0] & 68.0 \\
    \midrule
    \textbf{Overall}
      & \textbf{286/350 (81.7\%)} & \textbf{[77.3,\,85.4]}
      & \textbf{193/350 (55.1\%)} & \textbf{[49.9,\,60.3]}
      & \textbf{26.6} \\
    Restricted (4 templates)
      & 136/200 (68.0\%) & ---
      & 43/200 (21.5\%)  & ---
      & 46.5 \\
    \bottomrule
  \end{tabular}
\end{table*}

\section{Related Work} \label{sec:related}
We organize prior work along the three axes that together define the gap \textsc{GapFuzz} addresses---distributed-cluster targeting, concurrent injection, and a cross-plane oracle---and conclude with verification approaches that complement runtime divergence detection.

\noindent
\textbf{Single-controller SDN testing.} The first generation of SDN security testing focused on a single controller and its application interface. DELTA~\cite{lee2017delta} replays known attack scenarios and discovers new ones through fuzzing of input validation and application-layer abuses; BEADS~\cite{jero2017beads} fuzzes control-plane messages while staying within OpenFlow protocol constraints, exercising components beyond protocol parsers. Subsequent work extends along similar single-controller axes: model-based black-box testing~\cite{8107437}, broad-spectrum penetration assessment~\cite{LEE2020101720}, optimized fuzzing campaigns~\cite{chi2023vulnerability}, and intent-state-transition-guided fuzzing for intent-based networking~\cite{287272}. NICE~\cite{canini2012nice} pioneered the combination of model checking and symbolic execution on OpenFlow controller code to expose policy violations, and learning-guided fuzzing has recently been applied to stateful SDN controllers in single-instance deployments~\cite{ollando2026learning}. None of these tools exercises the temporal dynamics of distributed replication, which sit outside their threat model by construction.

\noindent
\textbf{Distributed SDN testing.} In the distributed setting, Ambusher~\cite{10534286} is a protocol-state-aware fuzzer for the East-West communication of SDN controller clusters. Its detection criteria are restricted to control-plane signals---cluster configuration changes, resource usage, application state---and it never queries the actual forwarding state of the switch; it therefore cannot expose a divergence between the cluster's authoritative state and the rule effectively applied to traffic. Empirical studies of consensus replication in distributed SDN~\cite{8470166,sakic2019response} characterize the latency-versus-consistency trade-off that creates the temporal window we exploit, but stop short of weaponizing it as a fuzzing axis. Race conditions in the SDN control plane have been demonstrated by~\cite{203884} and the more recent intent-based timing attacks of~\cite{10.1145/3658644.3670301}, both for single-controller settings.

\noindent
\textbf{Cross-plane validation.} AudiSDN~\cite{9681706,9155378} pioneered the systematic comparison of the controller's policy database against the flow table on the switch to detect deterministic inconsistencies caused by input-validation flaws and protocol-version mismatches. AudiSDN explicitly excludes from its scope transient inconsistencies that arise during normal operation (e.g., the window before a \texttt{BARRIER\_RESPONSE} is received) and evaluates only single-controller deployments---neither distributed consensus nor concurrent injections lie in its scope. CHIMERA~\cite{11023293} extends validation to the data plane through static concolic execution on P4 grammars, targeting parser-level and switch-software defects under a non-replicated controller assumption. Network state fuzzing~\cite{shukla2020consistent} also compares control-plane intent against data-plane state, but on a single, non-replicated control plane, and without driving concurrent injection. The data plane itself has been studied as a vulnerable surface in cloud deployments~\cite{thimmaraju2018taking}, motivating the use of a probe at the kernel datapath rather than the OpenFlow user-space view.

\begin{sloppypar}
\noindent
\textbf{Verification and complementary approaches.} Formal verification of SDN policies and snapshots---NetKAT~\cite{10.1145/2535838.2535862}, VeriFlow~\cite{10.1145/2342441.2342452}, abstractions for consistent network update~\cite{reitblatt2012abstractions}, and security enforcement kernels such as FortNOX~\cite{porras2012security}---verifies invariants of the network configuration but does not exercise runtime replication dynamics, and therefore complements rather than replaces concurrency-driven divergence detection. Distributed-system fuzzers in the Jepsen tradition~\cite{jepsen} and recent greybox extensions to distributed protocols~\cite{meng2023greybox} target consistency of replicated databases via application-level clients and instrumented runtimes, but neither category models the cross-plane contract specific to SDN: a fault that manifests as a discrepancy between cluster state and the action enforced on a packet at a switch.
\end{sloppypar}

\noindent
\textbf{Synthesis.} Table~\ref{tab:related-comparison} compares prior work along the three discriminating dimensions. No prior tool combines distributed-cluster targeting, concurrent injection, and a cross-plane oracle. The combination is precisely what is needed to surface the class of fault demonstrated in Section~\ref{subsec:poc} and characterized by \textsc{GapFuzz} in Section~\ref{sec:eval:results}.

\begin{table}[ht!]
\centering
\caption{Comparison of SDN testing approaches along three discriminating dimensions: \textit{Distributed SDN cluster} (evaluation on a multi-node controller cluster), \textit{Concurrent injection} (simultaneous contradictory requests through the controller interface), and \textit{Cross-plane oracle} (comparison between cluster state and the forwarding action effectively applied at the switch).}
\label{tab:related-comparison}
\begin{tabular}{lccc}
\toprule
Approach & \makecell{Distributed\\SDN cluster} & \makecell{Concurrent\\injection} & \makecell{Cross-plane\\oracle} \\
\midrule
BEADS~\cite{jero2017beads}              & \xmark & \xmark & \xmark \\
DELTA~\cite{lee2017delta}               & \xmark & \xmark & \xmark \\
NICE~\cite{canini2012nice}              & \xmark & \xmark & \xmark \\
Intender~\cite{287272}                  & \xmark & \xmark & \xmark \\
Ambusher~\cite{10534286}                & \cmark & \xmark & \xmark \\
AudiSDN~\cite{9681706,9155378}          & \xmark & \xmark & \cmark \\
CHIMERA~\cite{11023293}                 & \xmark & \xmark & \cmark \\
Net-state fuzzing~\cite{shukla2020consistent} & \xmark & \xmark & \cmark \\
\textbf{\textsc{GapFuzz} (our work)}    & \cmark & \cmark & \cmark \\
\bottomrule
\end{tabular}
\end{table}

\section{Threats to Validity}\label{sec:threats}
\begin{sloppypar}

\noindent
\textbf{Internal validity.} A divergence reported by \textsc{GapFuzz} could in principle reflect stale state from earlier runs rather than the current race. We mitigate this with a three-layer reset (cluster store, OpenFlow user-space table, kernel datapath megaflow cache) executed before every injection; without the third layer, megaflows survive control-plane cleanup and pollute later observations~\cite{pfaff2015design}, an artifact we encountered during the initial campaign and that motivated adding the cache flush. A second risk is that requests reaching the master directly bypass the asynchronous replication path of \S\ref{subsec:rca}; mastership-aware injection therefore targets only non-master nodes. The algorithm parameters $\Delta t_0$, $\varepsilon$, $\delta$, $\tau_\ell$, and $D_{\max}$ all influence detection sensitivity. We use the values reported in \S\ref{sec:eval:setup} and leave a sensitivity analysis to future work; the doubling-cap interpretation in \S\ref{sec:eval:results} explicitly acknowledges that $\Delta t_{\max} = 10.24$\,s is a budget bound, not a measurement.

\noindent
\textbf{External validity.} Our results come from a single configuration: three ONOS~2.7 nodes managing one Open~vSwitch over OpenFlow~1.3 inside a Mininet topology. We do not test other controllers (OpenDaylight, Ryu), other ONOS versions, hardware switches, or other OpenFlow versions. The seven templates used cover six of the fourteen match fields and four of the six actions in scope of the model in \S\ref{sec:model}; the uncovered combinations (\texttt{GROUP}, \texttt{PUSH\_VLAN}, \texttt{ICMP}, \texttt{UDP\_DST}, \texttt{SET\_FIELD} on L4 ports) may behave differently under the same fuzzing strategy. The kernel-datapath staleness we observe is specific to the Open~vSwitch revalidator~\cite{pfaff2015design}; switches with different reconciliation behavior may reconcile faster or under different triggers. We expect the underlying \emph{structure} of the fault---a commit-to-replicate gap producing a cross-plane divergence---to generalize across platforms that adopt asynchronous flow-state replication~\cite{8470166,sakic2019response,SMYTH20199}, but generalizing the empirical numbers remains an open question.

\noindent
\textbf{Construct validity.} Our data-plane oracle uses \texttt{ovs-appctl ofproto\allowbreak/trace}, which traces a simulated packet through the switch pipeline; we extract the datapath action from the \texttt{Datapath actions} line. The trace simulates rather than transmits a real packet, so the simulation may diverge from in-flight traffic under high load or when stateful pipeline elements depend on prior packets. The four-class classifier of \S\ref{sec:model} is defined at $t > T_2$; \texttt{CP\_DIVERGENT} and \texttt{FULL\_SPLIT} did not appear in our experiments because the cluster always converged within $T_2$ on our corpus, leaving only the upper row of the divergence matrix populated. A campaign that perturbs the East-West channel itself, or that injects on more than two non-master nodes simultaneously, may exercise the lower row.

\noindent
\textbf{Conclusion validity.} We run two campaigns: a full-mode campaign at $N=32$ (Algorithm~\ref{alg:gapfuzz} with Phase~1 + Phase~2; one campaign aborted before its last template due to a companion-process crash, leaving 223 attempts) and a Phase-1-only baseline at $N=50$ (350 attempts). Wilson 95\% confidence intervals~\cite{wilson1927probable} are narrowest when hit rates approach 0 or 1 and widest for race-bounded templates whose hit rates approach 50\% (\texttt{drop\_vs\_output} therefore exhibits the largest variability, as its trigger is timing-sensitive); per-template intervals appear in \S\ref{sec:eval:results}. We do not run BEADS, DELTA, or Ambusher on our setup; instead, the ablation in \S\ref{sec:eval:ablation} swaps \textsc{GapFuzz}'s kernel-datapath oracle for the OpenFlow user-space probe those tools rely on and runs the same workload through both, producing an upper bound on what a single-plane oracle would detect on this workload.

\end{sloppypar}

\section{Data Availability} 
All data, source code, templates, and run scripts required to reproduce the results of this paper are provided in our replication package, available at \url{https://github.com/mad975/GapFuzz}. The package includes the \textsc{GapFuzz} driver, the seven YAML templates of \S\ref{sec:eval:setup}, the Ryu companion exposing \texttt{ofproto/trace}, the three-layer reset scripts, and the raw per-run logs underlying every figure and table. Detailed documentation accompanies the package to facilitate independent reproduction.\label{sec:data}

\section{Conclusion}
\begin{sloppypar}
We presented \textsc{GapFuzz}, a stateful concurrency fuzzer that exposes cross-plane divergences in distributed SDN clusters: states in which the action a packet effectively receives at the data plane disagrees with what the cluster's authoritative state describes, even after the cluster has internally converged. \textsc{GapFuzz} couples a Northbound injector that targets two non-master nodes with a controlled inter-injection delay $\Delta t$ and a Differential Oracle that reads the kernel-datapath action via \texttt{ovs-appctl ofproto/trace}, and it characterizes each detected divergence on two orthogonal temporal axes through a two-phase timing search and a lifetime probe. Each verdict maps onto one of four mutually exclusive cross-plane state classes derived from a state model formalized against the ONOS~2.7 source.

On a three-node ONOS~2.7.0 cluster managing a single Open~vSwitch over OpenFlow~1.3, \textsc{GapFuzz} produces a divergent verdict in 81.7\% of attempts ($N=50$, Wilson 95\% CI: 77.3--85.4\%); all 286 divergences sit at the data plane, since the cluster nodes do not internally disagree under our two-non-master injection pattern. The temporal analysis separates one race-bounded template (\texttt{drop\_vs\_output}, $\Delta t_{\max} = 5$\,ms) from six that reach the doubling cap at 10.24\,s, and the lifetime probe shows that 99.4\% of full-mode divergences persist past $\tau_{\ell} = 30$\,s. Replacing the kernel-datapath probe with the OpenFlow user-space probe used by prior fuzzers drops detection from 81.7\% to 55.1\% on the full template set, and from 68.0\% to 21.5\% on the four templates where canonicalization does not force the verdict; the resulting 46.5-point gap is a lower bound on the cross-plane signal that controller- or OpenFlow-level oracles miss on this workload.

The evaluation has two acknowledged limits. It runs on a single cluster version and a single-switch topology, so generalization to other SDN stacks, other ONOS versions, and multi-switch topologies remains open. The canonical action set $\{\texttt{DROP}, \texttt{OUTPUT}(\text{port})\}$ used by the seed generator cannot represent SET\_FIELD- or POP\_VLAN-only actions, so three of the seven templates report \texttt{DP\_DIVERGENT} for an action-representation reason rather than a substantive cross-plane gap; lifting this restriction would extend the analysis to all seven templates and is the natural next step.
\end{sloppypar}

\section{Acknowledgments}
This work was supported by (1) the Luxembourg Ministry of Foreign and European Affairs through their Digital4Development (D4D) portfolio under project LuxWAyS and (2) the European Research Council (ERC) under the European Union’s Horizon 2020 research and innovation programme (Project NATURAL - grant agreement No 949014).

\bibliographystyle{ACM-Reference-Format}
\bibliography{sample-base}

@String{Computing = "Computing" }

@String{Computer = "{IEEE} Computer" }

@String{Springer = "Springer-Verlag" }

@inproceedings{10.1145/2486001.2486019,
author = {Jain, Sushant and Kumar, Alok and Mandal, Subhasree and Ong, Joon and Poutievski, Leon and Singh, Arjun and Venkata, Subbaiah and Wanderer, Jim and Zhou, Junlan and Zhu, Min and Zolla, Jon and H\"{o}lzle, Urs and Stuart, Stephen and Vahdat, Amin},
title = {B4: experience with a globally-deployed software defined wan},
year = {2013},
isbn = {9781450320566},
publisher = {Association for Computing Machinery},
address = {New York, NY, USA},
url = {https://doi.org/10.1145/2486001.2486019},
doi = {10.1145/2486001.2486019},
booktitle = {Proceedings of the ACM SIGCOMM 2013 Conference on SIGCOMM},
pages = {3–14},
numpages = {12},
location = {Hong Kong, China},
series = {SIGCOMM '13}
}

@article{10.1145/2534169.2486012,
author = {Hong, Chi-Yao and Kandula, Srikanth and Mahajan, Ratul and Zhang, Ming and Gill, Vijay and Nanduri, Mohan and Wattenhofer, Roger},
title = {Achieving high utilization with software-driven WAN},
year = {2013},
issue_date = {October 2013},
publisher = {Association for Computing Machinery},
address = {New York, NY, USA},
volume = {43},
number = {4},
issn = {0146-4833},
url = {https://doi.org/10.1145/2534169.2486012},
doi = {10.1145/2534169.2486012},
journal = {SIGCOMM Comput. Commun. Rev.},
month = aug,
pages = {15–26},
numpages = {12}
}

@inproceedings{10.1145/3243734.3243759,
author = {Ujcich, Benjamin E. and Jero, Samuel and Edmundson, Anne and Wang, Qi and Skowyra, Richard and Landry, James and Bates, Adam and Sanders, William H. and Nita-Rotaru, Cristina and Okhravi, Hamed},
title = {Cross-App Poisoning in Software-Defined Networking},
year = {2018},
isbn = {9781450356930},
publisher = {Association for Computing Machinery},
address = {New York, NY, USA},
url = {https://doi-org.proxy.bnl.lu/10.1145/3243734.3243759},
doi = {10.1145/3243734.3243759},
booktitle = {Proceedings of the 2018 ACM SIGSAC Conference on Computer and Communications Security},
pages = {648–663},
numpages = {16},
location = {Toronto, Canada},
series = {CCS '18}
}

@inproceedings{10.1145/2876019.2876024,
author = {Lee, Seungsoo and Yoon, Changhoon and Shin, Seungwon},
title = {The Smaller, the Shrewder: A Simple Malicious Application Can Kill an Entire SDN Environment},
year = {2016},
isbn = {9781450340786},
publisher = {Association for Computing Machinery},
address = {New York, NY, USA},
url = {https://doi-org.proxy.bnl.lu/10.1145/2876019.2876024},
doi = {10.1145/2876019.2876024},
booktitle = {Proceedings of the 2016 ACM International Workshop on Security in Software Defined Networks \& Network Function Virtualization},
pages = {23–28},
numpages = {6},
location = {New Orleans, Louisiana, USA},
series = {SDN-NFV Security '16}
}

@INPROCEEDINGS{10229058,
  author={Deng, Shuhua and Qing, Xian and Li, Xiaofan and Gao, Xing and Gao, Xieping},
  booktitle={IEEE INFOCOM 2023 - IEEE Conference on Computer Communications}, 
  title={SDN Application Backdoor: Disrupting the Service via Poisoning the Topology}, 
  year={2023},
  volume={},
  number={},
  pages={1-10},
  doi={10.1109/INFOCOM53939.2023.10229058}}

@ARTICLE{10332465,
  author={Deng, Shuhua and Chen, Lihui and Gao, Xieping},
  journal={IEEE Transactions on Network and Service Management}, 
  title={Manipulating Sensitive Match Fields to Poison Applications in SDN}, 
  year={2024},
  volume={21},
  number={2},
  pages={2413-2425},
  doi={10.1109/TNSM.2023.3337434}}

@article{10.1145/3453648,
author = {Rauf, Bilal and Abbas, Haider and Usman, Muhammad and Zia, Tanveer A. and Iqbal, Waseem and Abbas, Yawar and Afzal, Hammad},
title = {Application Threats to Exploit Northbound Interface Vulnerabilities in Software Defined Networks},
year = {2021},
issue_date = {July 2022},
publisher = {Association for Computing Machinery},
address = {New York, NY, USA},
volume = {54},
number = {6},
issn = {0360-0300},
url = {https://doi.org/10.1145/3453648},
doi = {10.1145/3453648},
journal = {ACM Comput. Surv.},
month = jul,
articleno = {121},
numpages = {36}
}

@inproceedings{jero2017beads,
  title={Beads: Automated attack discovery in openflow-based sdn systems},
  author={Jero, Samuel and Bu, Xiangyu and Nita-Rotaru, Cristina and Okhravi, Hamed and Skowyra, Richard and Fahmy, Sonia},
  booktitle={International Symposium on Research in Attacks, Intrusions, and Defenses},
  pages={311--333},
  year={2017},
  organization={Springer}
}

@INPROCEEDINGS{8107437,
  author={Yao, Jiangyuan and Wang, Zhiliang and Yin, Xia and Shi, Xingang and Li, Yahui and Li, Chongrong},
  booktitle={2017 IEEE 25th International Symposium on Modeling, Analysis, and Simulation of Computer and Telecommunication Systems (MASCOTS)}, 
  title={Testing Black-Box SDN Applications with Formal Behavior Models}, 
  year={2017},
  volume={},
  number={},
  pages={110-120},
  doi={10.1109/MASCOTS.2017.20}}

@INPROCEEDINGS{8526819,
  author={Lee, Chanhee and Yoon, Changhoon and Shin, Seungwon and Cha, Sang Kil},
  booktitle={2018 IEEE 26th International Conference on Network Protocols (ICNP)}, 
  title={INDAGO: A New Framework For Detecting Malicious SDN Applications}, 
  year={2018},
  volume={},
  number={},
  pages={220-230},
  doi={10.1109/ICNP.2018.00031}}

@ARTICLE{10534286,
  author={Kim, Jinwoo and Seo, Minjae and Marin, Eduard and Lee, Seungsoo and Nam, Jaehyun and Shin, Seungwon},
  journal={IEEE Transactions on Information Forensics and Security}, 
  title={Ambusher: Exploring the Security of Distributed SDN Controllers Through Protocol State Fuzzing}, 
  year={2024},
  volume={19},
  number={},
  pages={6264-6279},
  doi={10.1109/TIFS.2024.3402967}}

@article{LEE2020101720,
title = {A comprehensive security assessment framework for software-defined networks},
journal = {Computers \&  Security},
volume = {91},
pages = {101720},
year = {2020},
issn = {0167-4048},
doi = {https://doi.org/10.1016/j.cose.2020.101720},
url = {https://www.sciencedirect.com/science/article/pii/S0167404820300079},
author = {Seungsoo Lee and Jinwoo Kim and Seungwon Woo and Changhoon Yoon and Sandra Scott-Hayward and Vinod Yegneswaran and Phillip Porras and Seungwon Shin}
}

@inproceedings{chi2023vulnerability,
  title={A Vulnerability Detection Method for SDN with Optimized Fuzzing},
  author={Chi, Xiaofeng and Wang, Bingquan and Zhao, Jingling and Cui, Baojiang},
  booktitle={International Conference on Advanced Information Networking and Applications},
  pages={525--536},
  year={2023},
  organization={Springer}
}

@inproceedings {287272,
author = {Jiwon Kim and Benjamin E. Ujcich and Dave (Jing) Tian},
title = {Intender: Fuzzing {Intent-Based} Networking with {Intent-State} Transition Guidance},
booktitle = {32nd USENIX Security Symposium (USENIX Security 23)},
year = {2023},
isbn = {978-1-939133-37-3},
address = {Anaheim, CA},
pages = {4463--4480},
url = {https://www.usenix.org/conference/usenixsecurity23/presentation/kim-jiwon},
publisher = {USENIX Association},
month = aug
}

@ARTICLE{9681706,
  author={Lee, Seungsoo and Woo, Seungwon and Kim, Jinwoo and Nam, Jaehyun and Yegneswaran, Vinod and Porras, Phillip and Shin, Seungwon},
  journal={IEEE/ACM Transactions on Networking}, 
  title={A Framework for Policy Inconsistency Detection in Software-Defined Networks}, 
  year={2022},
  volume={30},
  number={3},
  pages={1410-1423},
  doi={10.1109/TNET.2022.3140824}}

@inproceedings{lee2017delta,
  title={Delta: A security assessment framework for software-defined networks.},
  author={Lee, Seungsoo and Yoon, Changhoon and Lee, Chanhee and Shin, Seungwon and Yegneswaran, Vinod and Porras, Phillip A},
  booktitle={NDSS},
  year={2017}
}

@INPROCEEDINGS{8023155,
  author={Ujcich, Benjamin E. and Thakore, Uttam and Sanders, William H.},
  booktitle={2017 47th Annual IEEE/IFIP International Conference on Dependable Systems and Networks (DSN)}, 
  title={ATTAIN: An Attack Injection Framework for Software-Defined Networking}, 
  year={2017},
  volume={},
  number={},
  pages={567-578},
  doi={10.1109/DSN.2017.59}}

@INPROCEEDINGS{11023293,
  author={Kim, Jiwon and Tian, Dave Jing and Ujcich, Benjamin E.},
  booktitle={2025 IEEE Symposium on Security and Privacy (SP)}, 
  title={Chimera: Fuzzing P4 Network Infrastructure for Multi-Plane Bug Detection and Vulnerability Discovery}, 
  year={2025},
  volume={},
  number={},
  pages={3088-3106},
  doi={10.1109/SP61157.2025.00194}}

@article{liang2018fuzzing,
  title={Fuzzing: State of the art},
  author={Liang, Hongliang and Pei, Xiaoxiao and Jia, Xiaodong and Shen, Wuwei and Zhang, Jian},
  journal={IEEE Transactions on Reliability},
  volume={67},
  number={3},
  pages={1199--1218},
  year={2018},
  publisher={IEEE}
}

@inproceedings {203884,
author = {Lei Xu and Jeff Huang and Sungmin Hong and Jialong Zhang and Guofei Gu},
title = {Attacking the Brain: Races in the {SDN} Control Plane},
booktitle = {26th USENIX Security Symposium (USENIX Security 17)},
year = {2017},
isbn = {978-1-931971-40-9},
address = {Vancouver, BC},
pages = {451--468},
url = {https://www.usenix.org/conference/usenixsecurity17/technical-sessions/presentation/xu-lei},
publisher = {USENIX Association},
month = aug
}

@inproceedings{10.1145/3658644.3670301,
author = {Weintraub, Ben and Kim, Jiwon and Tao, Ran and Nita-Rotaru, Cristina and Okhravi, Hamed and Tian, Dave (Jing) and Ujcich, Benjamin E.},
title = {Exploiting Temporal Vulnerabilities for Unauthorized Access in Intent-based Networking},
year = {2024},
isbn = {9798400706363},
publisher = {Association for Computing Machinery},
address = {New York, NY, USA},
url = {https://doi-org.proxy.bnl.lu/10.1145/3658644.3670301},
doi = {10.1145/3658644.3670301},
booktitle = {Proceedings of the 2024 on ACM SIGSAC Conference on Computer and Communications Security},
pages = {3630–3644},
numpages = {15},
location = {Salt Lake City, UT, USA},
series = {CCS '24}
}

@inproceedings{musuvathi2008finding,
  title={Finding and Reproducing Heisenbugs in Concurrent Programs.},
  author={Musuvathi, Madanlal and Qadeer, Shaz and Ball, Thomas and Basler, Gerard and Nainar, Piramanayagam Arumuga and Neamtiu, Iulian},
  booktitle={OSDI},
  volume={8},
  number={2008},
  year={2008}
}

@ARTICLE{8470166,
  author={Sakic, Ermin and Kellerer, Wolfgang},
  journal={IEEE Journal on Selected Areas in Communications}, 
  title={Impact of Adaptive Consistency on Distributed SDN Applications: An Empirical Study}, 
  year={2018},
  volume={36},
  number={12},
  pages={2702-2715},
  doi={10.1109/JSAC.2018.2871309},
  ISSN={1558-0008},
  month={Dec},}

@inproceedings{10.1145/2342441.2342443,
author = {Levin, Dan and Wundsam, Andreas and Heller, Brandon and Handigol, Nikhil and Feldmann, Anja},
title = {Logically centralized? state distribution trade-offs in software defined networks},
year = {2012},
isbn = {9781450314770},
publisher = {Association for Computing Machinery},
address = {New York, NY, USA},
url = {https://doi.org/10.1145/2342441.2342443},
doi = {10.1145/2342441.2342443},
booktitle = {Proceedings of the First Workshop on Hot Topics in Software Defined Networks},
pages = {1–6},
numpages = {6},
location = {Helsinki, Finland},
series = {HotSDN '12}
}

@ARTICLE{6994333,
  author={Kreutz, Diego and Ramos, Fernando M. V. and Veríssimo, Paulo Esteves and Rothenberg, Christian Esteve and Azodolmolky, Siamak and Uhlig, Steve},
  journal={Proceedings of the IEEE}, 
  title={Software-Defined Networking: A Comprehensive Survey}, 
  year={2015},
  volume={103},
  number={1},
  pages={14-76},
  doi={10.1109/JPROC.2014.2371999}}

@ARTICLE{8187644,
  author={Bannour, Fetia and Souihi, Sami and Mellouk, Abdelhamid},
  journal={IEEE Communications Surveys \& Tutorials}, 
  title={Distributed SDN Control: Survey, Taxonomy, and Challenges}, 
  year={2018},
  volume={20},
  number={1},
  pages={333-354},
  doi={10.1109/COMST.2017.2782482}}

@inproceedings{10.1145/2620728.2620744,
author = {Berde, Pankaj and Gerola, Matteo and Hart, Jonathan and Higuchi, Yuta and Kobayashi, Masayoshi and Koide, Toshio and Lantz, Bob and O'Connor, Brian and Radoslavov, Pavlin and Snow, William and Parulkar, Guru},
title = {ONOS: towards an open, distributed SDN OS},
year = {2014},
isbn = {9781450329897},
publisher = {Association for Computing Machinery},
address = {New York, NY, USA},
url = {https://doi.org/10.1145/2620728.2620744},
doi = {10.1145/2620728.2620744},
booktitle = {Proceedings of the Third Workshop on Hot Topics in Software Defined Networking},
pages = {1–6},
numpages = {6},
location = {Chicago, Illinois, USA},
series = {HotSDN '14}
}

@ARTICLE{6127847,
  author={Abadi, Daniel},
  journal={Computer}, 
  title={Consistency Tradeoffs in Modern Distributed Database System Design: CAP is Only Part of the Story}, 
  year={2012},
  volume={45},
  number={2},
  pages={37-42},
  doi={10.1109/MC.2012.33}}

@article{SMYTH20199,
title = {Attacking distributed software-defined networks by leveraging network state consistency},
journal = {Computer Networks},
volume = {156},
pages = {9-19},
year = {2019},
issn = {1389-1286},
doi = {https://doi.org/10.1016/j.comnet.2019.02.020},
url = {https://www.sciencedirect.com/science/article/pii/S138912861830940X},
author = {Dylan Smyth and Donna O’Shea and Victor Cionca and Sean McSweeney}
}

@incollection{lamport2019part,
  title={The part-time parliament},
  author={Lamport, Leslie},
  booktitle={Concurrency: the works of Leslie Lamport},
  pages={277--317},
  year={2019}
}

@inproceedings{ongaro2014search,
  title={In search of an understandable consensus algorithm},
  author={Ongaro, Diego and Ousterhout, John},
  booktitle={2014 USENIX annual technical conference (USENIX ATC 14)},
  pages={305--319},
  year={2014}
}

@misc{jepsen,
  author = {Kingsbury, Kyle},
  title = {Jepsen: Distributed Systems Safety Research},
  year = {2013},
  howpublished = {\url{https://jepsen.io/}},
  note = {Accessed: 2026-05-02}
}

@inproceedings{10.1145/2342441.2342452,
author = {Khurshid, Ahmed and Zhou, Wenxuan and Caesar, Matthew and Godfrey, P. Brighten},
title = {VeriFlow: verifying network-wide invariants in real time},
year = {2012},
isbn = {9781450314770},
publisher = {Association for Computing Machinery},
address = {New York, NY, USA},
url = {https://doi-org.proxy.bnl.lu/10.1145/2342441.2342452},
doi = {10.1145/2342441.2342452},
booktitle = {Proceedings of the First Workshop on Hot Topics in Software Defined Networks},
pages = {49–54},
numpages = {6},
location = {Helsinki, Finland},
series = {HotSDN '12}
}

@inproceedings{10.1145/2535838.2535862,
author = {Anderson, Carolyn Jane and Foster, Nate and Guha, Arjun and Jeannin, Jean-Baptiste and Kozen, Dexter and Schlesinger, Cole and Walker, David},
title = {NetKAT: semantic foundations for networks},
year = {2014},
isbn = {9781450325448},
publisher = {Association for Computing Machinery},
address = {New York, NY, USA},
url = {https://doi-org.proxy.bnl.lu/10.1145/2535838.2535862},
doi = {10.1145/2535838.2535862},
booktitle = {Proceedings of the 41st ACM SIGPLAN-SIGACT Symposium on Principles of Programming Languages},
pages = {113–126},
numpages = {14},
location = {San Diego, California, USA},
series = {POPL '14}
}

@article{wilson1927probable,
  title={Probable inference, the law of succession, and statistical inference},
  author={Wilson, Edwin B},
  journal={Journal of the American Statistical Association},
  volume={22},
  number={158},
  pages={209--212},
  year={1927},
  publisher={Taylor \& Francis}
}

@inproceedings{canini2012nice,
  title     = {A {NICE} Way to Test {OpenFlow} Applications},
  author    = {Canini, Marco and Venzano, Daniele and Peresíni, Peter and Kosti{\'c}, Dejan and Rexford, Jennifer},
  booktitle = {9th USENIX Symposium on Networked Systems Design and Implementation (NSDI 12)},
  year      = {2012},
  pages     = {127--140},
  address   = {San Jose, CA},
  publisher = {USENIX Association}
}

@inproceedings{pfaff2015design,
  title     = {The Design and Implementation of {Open vSwitch}},
  author    = {Pfaff, Ben and Pettit, Justin and Koponen, Teemu and Jackson, Ethan and Zhou, Andy and Rajahalme, Jarno and Gross, Jesse and Wang, Alex and Stringer, Joe and Shelar, Pravin and Amidon, Keith and Casado, Mart{\'i}n},
  booktitle = {12th USENIX Symposium on Networked Systems Design and Implementation (NSDI 15)},
  year      = {2015},
  pages     = {117--130},
  address   = {Oakland, CA},
  publisher = {USENIX Association}
}

@article{mckeown2008openflow,
  title    = {{OpenFlow}: Enabling Innovation in Campus Networks},
  author   = {McKeown, Nick and Anderson, Tom and Balakrishnan, Hari and Parulkar, Guru and Peterson, Larry and Rexford, Jennifer and Shenker, Scott and Turner, Jonathan},
  journal  = {ACM SIGCOMM Computer Communication Review},
  volume   = {38},
  number   = {2},
  pages    = {69--74},
  year     = {2008},
  doi      = {10.1145/1355734.1355746}
}

@article{ollando2026learning,
  title   = {Learning-Guided Fuzzing for Testing Stateful {SDN} Controllers},
  author  = {Ollando, Rapha{\"e}l and Shin, Seung Yeob and Briand, Lionel C.},
  journal = {ACM Transactions on Software Engineering and Methodology},
  year    = {2026},
  doi     = {10.1145/3733717}
}

@article{shukla2020consistent,
  title   = {Towards Consistent {SDNs} Through Network State Fuzzing},
  author  = {Shukla, Apoorv and Saidi, S. Jawad and Schmid, Stefan and Canini, Marco and Zinner, Thomas and Feldmann, Anja},
  journal = {IEEE Transactions on Network and Service Management},
  year    = {2020},
  volume  = {17},
  number  = {2},
  pages   = {1156--1169},
  doi     = {10.1109/TNSM.2019.2955790}
}

@inproceedings{reitblatt2012abstractions,
  title     = {Abstractions for Network Update},
  author    = {Reitblatt, Mark and Foster, Nate and Rexford, Jennifer and Schlesinger, Cole and Walker, David},
  booktitle = {Proceedings of the ACM SIGCOMM 2012 Conference},
  year      = {2012},
  pages     = {323--334},
  doi       = {10.1145/2342356.2342427}
}

@inproceedings{meng2023greybox,
  title     = {Greybox Fuzzing of Distributed Systems},
  author    = {Meng, Ruijie and P{\^i}rlea, George and Roychoudhury, Abhik and Sergey, Ilya},
  booktitle = {Proceedings of the 2023 ACM SIGSAC Conference on Computer and Communications Security (CCS '23)},
  year      = {2023},
  pages     = {1615--1629},
  doi       = {10.1145/3576915.3623097}
}

@article{sakic2019response,
  title   = {Response Time and Availability Study of {RAFT} Consensus in Distributed {SDN} Control Plane},
  author  = {Sakic, Ermin and Kellerer, Wolfgang},
  journal = {IEEE Transactions on Network and Service Management},
  volume  = {15},
  number  = {1},
  pages   = {304--318},
  year    = {2018},
  doi     = {10.1109/TNSM.2017.2775061}
}

@inproceedings{porras2012security,
  title     = {A Security Enforcement Kernel for {OpenFlow} Networks},
  author    = {Porras, Philip and Shin, Seungwon and Yegneswaran, Vinod and Fong, Martin and Tyson, Mabry and Gu, Guofei},
  booktitle = {Proceedings of the First Workshop on Hot Topics in Software Defined Networks (HotSDN '12)},
  year      = {2012},
  pages     = {121--126},
  doi       = {10.1145/2342441.2342466}
}

@inproceedings{thimmaraju2018taking,
  title     = {Taking Control of {SDN}-based Cloud Systems via the Data Plane},
  author    = {Thimmaraju, Kashyap and Shastry, Bhargava and Fiebig, Tobias and Hetzelt, Felicitas and Seifert, Jean-Pierre and Feldmann, Anja and Schmid, Stefan},
  booktitle = {Proceedings of the Symposium on SDN Research (SOSR '18)},
  year      = {2018},
  pages     = {1--15},
  doi       = {10.1145/3185467.3185468}
}

@inproceedings{kreutz2013towards,
  title     = {Towards Secure and Dependable Software-Defined Networks},
  author    = {Kreutz, Diego and Ramos, Fernando M.~V. and Verissimo, Paulo},
  booktitle = {Proceedings of the Second ACM SIGCOMM Workshop on Hot Topics in Software Defined Networking (HotSDN '13)},
  year      = {2013},
  pages     = {55--60},
  doi       = {10.1145/2491185.2491199}
}

@article{diouf2025software,
  title={Software security in software-defined networking: A systematic literature review},
  author={Diouf, Moustapha Awwalou and Ouya, Samuel and Klein, Jacques and Bissyand{\'e}, Tegawend{\'e} F},
  journal={arXiv preprint arXiv:2502.13828},
  year={2025}
}

@inproceedings{9155378,
    author = {Lee, Seungsoo and Woo, Seungwon and Kim, Jinwoo and Yegneswaran, Vinod and Porras, Phillip and Shin, Seungwon},
    booktitle = {IEEE INFOCOM 2020 - IEEE Conference on Computer Communications},
    doi = {10.1109/INFOCOM41043.2020.9155378},
    issn = {2641-9874},
    month = {July},
    number = {},
    pages = {1788-1797},
    title = {AudiSDN: Automated Detection of Network Policy Inconsistencies in Software-Defined Networks},
    volume = {},
    year = {2020}
}

\appendix

\end{document}